\documentclass[gsifonts,twoside,12pt]{gsipaper}
\usepackage{a4wide,gsiindex,helvet}
\usepackage[english]{babel}
\usepackage{amsfonts,amssymb,verbatim,graphicx,subfigure,float,epsfig,
psfrag,mathrsfs}
\usepackage{bm}

\usepackage[utf8]{inputenc}
\usepackage{pstricks}
\usepackage{pst-plot}

\def\slashchar#1{\setbox0=\hbox{$#1$}
   \dimen0=\wd0 \setbox1=\hbox{/} \dimen1=\wd1
   \ifdim\dimen0>\dimen1 \rlap{\hbox to \dimen0{\hfil/\hfil}} #1
   \else  \rlap{\hbox to \dimen1{\hfil$#1$\hfil}} / \fi}

\begin{document}

\title{QCD-Thermodynamics using 5-dim Gravity\\ 
}

\date{\today}

\abstract{ We calculate the critical temperature and free energy of
the gluon plasma using the dilaton potential~\cite{Galow:2009kw} in
the gravity theory of AdS/QCD. The finite temperature observables are
calculated in two ways: first, from the Page-Hawking computation of
the free energy, and secondly using the Bekenstein-Hawking
proportionality of the entropy with the area of the horizon.
Renormalization is well defined, because the $T=0$ theory has
asymptotic freedom. We further investigate the change of the critical
temperature with the number of flavours induced by the change of the
running coupling constant in the quenched theory.  The finite
temperature behaviour of the speed of sound, spatial string tension
and vacuum expectation value of the Polyakov loop follow from the
corresponding string theory in $AdS_5$.}

\author{E. Meg\'{\i}as$^a$, H.J. Pirner$^{ab}$ and  K. Veschgini$^a$}

\address{$\phantom{!}^{a}$ Institute for Theoretical Physics, University of Heidelberg, Germany\\
$\phantom{!}^{b}$ Max Planck Institute for Nuclear Physics,
Heidelberg, Germany}

\maketitle

\section{INTRODUCTION}
\label{sec:intro}

In the bottom-up approach of AdS/QCD important properties of pure glue
QCD are encoded in a phenomenological gravity theory through the
introduction of a dilaton potential.  The fifth dimension plays the
role of an inverse energy scale, which necessitates that the dilaton
is not constant, but runs with this scale. In a previous
paper~\cite{Galow:2009kw} we have fixed the ultraviolet behaviour of
the potential to the two loop beta function of QCD and parametrized
the infrared part in such a way that the heavy $Q \bar Q$ potential is
reproduced. In five dimensions the string connecting the $Q$ and $
\bar Q$ hangs into the bulk fifth dimension and thereby is sensitive
to the geometry of the five dimensional space. An obvious next step is
to consider properties of QCD in other simple environment where the
geometry of the five dimensional space changes in a controlled manner.

Finite temperature properties of QCD are on top of the list, since
they mostly concern spatially homogeneous systems, where the equations
of motion are still simple to solve. There has been quite some
understanding of the deconfinement transition in 4-dimensions on the
basis of strings in strong coupling lattice QCD. It comes from a
roughening of the strings due the entropic enhancement of
configurations with long wiggly strings. The high temperature phase,
however, is not understood in a picture where the underlying degrees
of freedom are strings at low temperature. Indeed the old Hagedorn
picture is limited to temperatures below the phase transition. Above
the critical temperature $T_c$ spatio-temporal Wilson loops are no
longer suppressed due to their area, the string tension goes to zero
and therefore strings have seized to live in the plasma. There is a
remnant of the low temperature theory in the behaviour of purely
spatial Wilson loops, but this effect is not very strong at high
temperatures. In Refs.~\cite{Antonov:2008kb} an analysis of the
contribution due to these spatial surfaces has been made in 4-
dimensions. There have been various attempts to analytically continue effective 4-dimensional string theories to the deconfinement phase, see e.g. Refs.~\cite{Polchinski:1992ty,Diamantini:2002rp,Kogan:2002au}, however without any phenomenological application so far.

The duality of string theory with 5-dimensional gravity can help. In
conformal $AdS_5$ the metric is well known. It has a horizon in the
bulk space at $r_T= \frac{\pi \ell^2}{ \beta}$ where $\beta=1/T$, the
inverse of temperature, and $\ell$ is the radius of the AdS-space.
Conformal solutions for entropy scale like $s \propto T^3$, since the
3-dimensional area of the horizon is given as $A \propto 2 \pi^2
r_T^3$. Promising solutions of this conformal theory have been
proposed to the problem of viscosity $\eta$~\cite{Policastro:2001yc}
with a small constant value for $\eta/s$. Top to bottom approaches
based on the conformal SYM with fermions have been investigating the
chiral phase transition and problems at finite density
\cite{Erdmenger:2007cm,Mateos:2006nu}.

Indeed there may be a window with the plasma not close to the phase
transition and not yet perturbative, where this conformal theory
mimics truthfully reality. With solutions at $T=0$ at
hand~\cite{Galow:2009kw,Pirner:2009gr} which break conformal symmetry
and give confinement, it is challenging to investigate the $T \neq 0$
sector of the theory for all temperatures.  Important progress has
been made in this direction by two groups
\cite{Gursoy:2008za,Alanen:2009xs}.

In this paper we follow this general approach, and investigate the
equation of state with two different methods. On the one hand we use
the Hawking-Page formalism \cite{Hawking:1982dh} to derive the
renormalized free energy from the action. On the other hand we
directly calculate the entropy from the Bekenstein-Hawking formula
\cite{Bekenstein:1973ur}. These two approaches are assumed to give
identical results as long as the black hole is treated entirely
classical. Renormalization is well defined, because the $T=0$ theory
has asymptotic freedom. We further investigate the change of the
critical temperature with the number of flavours induced by the change
of the running coupling constant.  In a renormalization group
framework interesting behaviour of $T_c$ has been demonstrated
recently \cite{Braun:2009ns}. Finally using the string theory we can
investigate the finite temperature behaviour of several thermodynamic
quantities, like the speed of sound, spatial string tension and vacuum
expectation value of the Polyakov loop. Comparison with lattice data
is made.
 
The outline of the paper is as follows: In Section 2 we give an
overview of the 5-d gravity action and specify the dilaton potential
which is used. In Section~\ref{sec:einsteinequations} we will
reproduce the zero temperature results focussing on the behaviour of
the metric and the coupling near the boundary $z \to
0$. Section~\ref{sec:bhsol} gives the finite temperature calculation
emphasizing the similarities and differences of the two solutions at
the boundary.  Section~\ref{sec:freeenergyEH} is devoted to deduce the
thermodynamics of the plasma using the Page-Hawking approach. In
Section~\ref{sec:Bekenstein_Hawking} we discuss the Bekenstein-Hawking
approach to compute the free energy, and compare with the method of
previous section. We also discuss the consequences of the flavour
dependence of the running coupling. In
Section~\ref{sec:thermodynamics} we compute the thermal behaviour of
speed of sound, spatial string tension and vacuum expectation value of
the Polyakov loop, and its comparison with available lattice
data. Finally, Section~\ref{sec:discussion} gives a final discussion
and our conclusions. In Appendix~A we discuss in details the procedure
to solve numerically the Einstein equations at finite temperature. In
Appendix~B we show technical details for the analytical computation of
thermodynamic quantities in the ultraviolet regime.

\section{5-d gravity action}

In the large $N_c$ limit, we assume a five
dimensional gravity-dilaton model, with the action

\begin{equation}
{\cal S}=\frac{1}{16 \pi G_5}\int d^5x \sqrt{G}\left(R-\frac{4}{3}\partial
_{\mu }\phi \partial ^{\mu }\phi -V(\phi )\right) -\frac{1}{8 \pi G_5}
\int _{\partial M} d^4x \sqrt{H} K  \,. \label{eq:action5D}
\end{equation}

The last term is the Gibbons-Hawking term, where the integration is
evaluated at the boundary $\partial M$ of the five dimensional space
given by $z =0$. The induced metric on the surface is denoted by $H$.
An infinitesimal distance $z=\epsilon $ to the boundary is used to
regularize our expressions.  Later we will take the limit \(\epsilon
\to 0\). The metric $G_{\mu\nu}$ is taken in the Einstein frame, and
the extrinsic curvature or second fundamental form of the boundary,
$K_{\mu\nu}$, is evaluated with the help of the normal $n^{\rho }$ at
the boundary $\partial M$:

\begin{eqnarray}
K_{\mu \nu}&=&  -\nabla_\mu n_\nu  = \frac{1}{2}n_{\rho } H^{\rho \sigma }\partial _{\sigma }H_{\mu \nu } \,, \\
K          &=& H^{\mu \nu}K_{\mu \nu}  \,.
\end{eqnarray}

In Ref.~\cite{Galow:2009kw} we extrapolated the $\beta$-function of
QCD to the infrared with a parametrization which was consistent with
asymptotic freedom and the heavy ${\bar q}q$ potential at zero
temperature. This parametrization takes the form

\begin{eqnarray}
  \beta(\alpha) &=& -b_2\alpha + \bigg[b_2\alpha  +
\left(\frac{b_2}{\bar{\alpha}}-\beta_0\right)\alpha^2 +
\left(\frac{b_2}{2\bar{\alpha}^2} - \frac{\beta_0}{\bar{\alpha}}-\beta_1 \right)\alpha^3 \bigg] e^{-\alpha/\bar{\alpha}}\,,
\label{eq:betaparam}
\end{eqnarray}
where 
\begin{equation}
\alpha(z) = e^{\phi(z)} \,,
\label{eq:alpha}
\end{equation}
is the running coupling. With this $\beta$-function we could obtain the dilaton potential in AdS/QCD as a function of the running coupling constant:

\begin{eqnarray}
  V(\alpha) &=& -\frac{12}{\ell^2}\left(1 - \left(\frac{\beta(\alpha)}{3
\alpha}\right)^2\right)
\left(\frac{\alpha}{\bar{\alpha}}\right)^\frac{8 b_2}{9} \nonumber \\
{} && \quad\cdot \mathrm{Exp}\left[\frac{4}{9} \left( (2\gamma-3)b_2 +
4 \beta_0\bar{\alpha} + 2 \beta_1\bar{\alpha}^2 \right)\right]
\nonumber \\ {} && \quad\cdot \mathrm{Exp} \left[ \frac{4}{9}
e^{-\frac{\alpha}{\bar{\alpha}}} \left( 3 b_2 - 4 \beta_0 \bar{\alpha}
- 2 \beta_1\bar{\alpha}^2 + (\frac{b_2}{\bar{\alpha}} -2 \beta_0 - 2
\beta_1 \bar{\alpha})\alpha\right) \right] \nonumber \\ {} &&
\quad\cdot \mathrm{Exp}\left( \frac{8 b_2}{9}\cdot
\mathrm{E_1}\left[\frac{\alpha}{\bar{\alpha}}\right]
\right)\,, \label{eq:Vanalytic}
\end{eqnarray}
where $E_1$ is the exponential integral function. For $\alpha <\bar{\alpha}$ the potential is strictly determined by the perturbative $\beta$-function,

\begin{eqnarray}
\beta(\alpha )&=&-\beta_0\alpha^2 - \beta_1\alpha^3 -\beta_2\alpha^4 +  \dots \\
\beta_0&=&\frac{1}{2\pi}\left(\frac{11}{3}N_c-\frac{2}{3}N_f\right) \label{eq:b0}\\
\beta_1&=&\frac{1}{8\pi^2}\left(\frac{34}{3} N_c^2-\left(\frac{13}{3}
N_c-\frac{1}{N_c}\right)N_f\right)  \label{eq:b1}\,.
\end{eqnarray}
The parameters $\beta_0$ and $\beta_1$ are universal, i.e. they are
regularization scheme independent.  For $\alpha >\bar{\alpha}$ the
$\beta$-function behaves linearly, $\beta(\alpha) \simeq - b_2 \alpha$,
and the potential is characterized by the non-perturbative constants
$-b_2 $ and $\bar{\alpha}$.  The parametrized coefficient $\beta_2$
has the form
\begin{equation}
\beta_2= \frac{b_2-3 \beta_0 \bar \alpha - 6 \beta_1 \bar \alpha^2}{6 \bar \alpha^3} \,.
\end{equation}

From the heavy $\bar{q}q$ potential at zero temperature one gets the optimum
values~\cite{Galow:2009kw}

\begin{equation}
b_2 = 2.3  \,, \qquad \bar\alpha = 0.45 \,,\qquad \ell   = 4.389 \, \textrm{GeV}^{-1} \,. \label{eq:b2alpha}
\end{equation}

The potential itself approaches the conformal limit $ V =-12/\ell^2$  
for $\alpha \to 0$. Based on this action we will investigate the thermodynamics of QCD.

\begin{figure}[tbp]
\begin{center}
\epsfig{figure=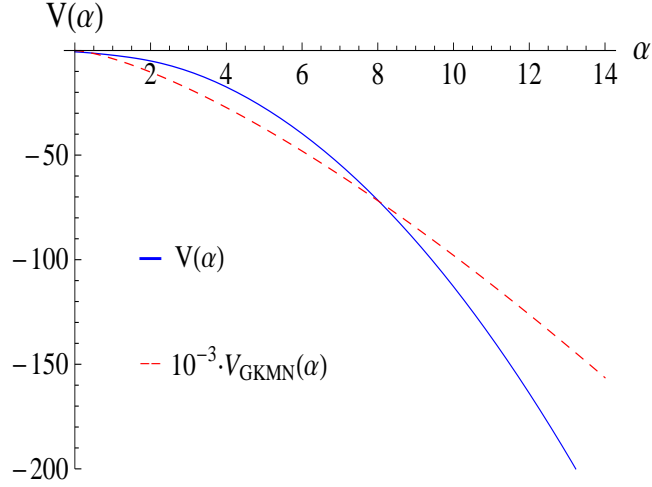,height=6.5cm,width=8.5cm}
\end{center}
\caption{Dilaton potential $V$ as a function of the running coupling $\alpha$. The full (blue) line shows the dilaton potential $V$ given by Eq.~(\ref{eq:Vanalytic}) with parameters Eqs.~(\ref{eq:b0}), (\ref{eq:b1}) and (\ref{eq:b2alpha}). For comparison, we show as a dashed (red) line the dilaton potential $V_{GKMN}$ given in Ref.~\cite{Gursoy:2009jd} multiplied by a factor $10^{-3}$.}
\label{fig:Valpha}
\end{figure}

Our parametrization given by Eq.~(\ref{eq:betaparam}) is simple and
more intuitive than that of Ref.~\cite{Gursoy:2009jd}. It allows for
analytical computations in many cases, e.g. we can derive analytically
the dilaton potential Eq.~(\ref{eq:Vanalytic}) from the
$\beta$-funtion Eq.~(\ref{eq:betaparam}). The dilaton potential of
Ref.~\cite{Gursoy:2009jd} is fine tuned and the role played by their
parameters are not so obvious. We plot in Fig.~\ref{fig:Valpha} the
dilaton potential of Eq.~(\ref{eq:Vanalytic}) which we use in this
work, and compare it with the one proposed in
Ref.~\cite{Gursoy:2009jd} which we call $V_{GKMN}$. We use in both
cases the same value of $\ell$ given by Eq.~(\ref{eq:b2alpha}).
$V_{GKMN}(\alpha)$ is of the order of $10^4\, \textrm{GeV}^2$ in the regime of
our interest, $\alpha \approx 0.3$, in contrast to the value $\sim 1
\, \textrm{GeV}^2$ given by our potential $V(\alpha)$. Both potentials
share the same ultraviolet behaviour, but they differ in the
infrared. Since the scheme dependent coefficient $\beta_2$ in
Ref.~\cite{Gursoy:2009jd} is very much larger than ours, the running
of the coupling of Ref.~\cite{Gursoy:2009jd} deviates already for
extremely small values from the regime dictated by the leading
coefficients $\beta_0$ and $\beta_1$ and then it depends
entirely on the ``infrared'' parametrisation in the regime of our
interest. In the scheme of Ref.~\cite{Gursoy:2009jd} the values of
$\alpha$ sampling the dilaton potential $V(\alpha)$ are much smaller
over the whole range of temperatures. At e.g. $T= 5T_c$
$\alpha(Ref.~\cite{Gursoy:2009jd})=0.00095$ or more than 100-times
smaller than the usual running coupling in $\overline{\textrm MS}$ scheme
$\alpha(\pi T)=0.11$. This explains the large difference between the
dilaton potentials when they are plotted as a function of the same
$\alpha$.  Unfortunately no mapping between
$\alpha(Ref.~\cite{Gursoy:2009jd})$ and $\alpha$ in our
$\overline{\textrm MS}$- scheme is known. A further discussion will be presented in Section~\ref{sec:Bekenstein_Hawking}.

\section{Thermal gas solution of the Einstein Equations }
\label{sec:einsteinequations}

The equations of motion corresponding to the five dimensional gravity-dilaton action, Eq.~(\ref{eq:action5D}), are given by

\begin{equation}
E_{\mu\nu}=T_{\mu\nu} \,, \label{eq:movmodel}
\end{equation}
with
\begin{eqnarray}
E_{\mu\nu}&=&R_{\mu \nu }-\frac{1}{2}R G_{\mu \nu } \,,\\   
T_{\mu\nu}&=&
\frac{4}{3}\partial _{\sigma }\partial _{\nu }\phi -\frac{1}{2}G_{\mu \nu }\left(\frac{4}{3}
\partial _{\sigma}\phi \partial ^{\sigma }\phi +V(\phi )\right) \,.
\end{eqnarray}
The left hand side of Eq.~(\ref{eq:movmodel}) is the Einstein tensor $E_{\mu \nu }$ 
while the right hand side is the energy momentum tensor $T_{\mu \nu }$. The {\it Thermal Gas} solution preserves spatial rotational invariance and it has
a metric similar to the zero temperature solution in Euclidean space:

\begin{eqnarray}
G_{\mu\nu} &=& b^2_0(z) \cdot \textrm{diag}\left(1,1,1,1,1\right) \,, \label{eq:Gth}  \\
ds^2       &=& b^2_0(z)\left(d\tau^2+dx_kdx^k+dz^2\right) \,, \label{eq:thermalmetric}
\end{eqnarray}
where \(z\in (0,\infty )\) is the bulk coordinate in the fifth
dimension, and the imaginary time coordinate $\tau$ is periodic
$(\tau \to \tau + \beta)$ with period $\beta=1/T$, the inverse of
temperature.  This solution exists at all temperatures.  Under the assumption that the energy scale is proportional to $b_0$, i.e. $E = \Lambda_E \, b_0$, the $\beta$-function writes 
\begin{equation}
\beta(\alpha) = b_0 \frac{d\alpha}{db_0}\,.
\end{equation}
There will be a second solution at finite temperature, the {\it Black
Hole} solution, with a horizon in the bulk coordinate which
characterizes the gluon plasma, and which we will discuss in the next
section.  After computing the Einstein tensor $E_{\mu\nu}$ and the energy momentum tensor $T_{\mu\nu}$ in terms of $b_0(z)$ and its derivatives, one ends up with three equations which determine the thermal gas solution:

\begin{eqnarray}
&&W'_0 = \frac{16}{9} b_0 W_0^2 +\frac{3}{4} b_0 V \,, \label{eq:em01}\\
&&b'_0 = -\frac{4}{9} b_0^2 W_0  \,, \label{eq:em02} \\
&&\alpha'_0 = \alpha_0 \sqrt{b_0 W'_0} \,. \label{eq:em03}
\end{eqnarray}

In the following text $(\;{}^\prime \;)$ stands for derivative with
respect to the $z$ coordinate.  In order to reduce the equations of
motion to first order equations, we have introduced the superpotential
defined as
\begin{equation}
W_0(z) \equiv -\frac{9}{4} \frac{b_0'(z)}{b_0^2(z)} \,. \label{eq:defW0} 
\end{equation}

This definition agrees with the one given in
Ref.~\cite{Galow:2009kw}.  The functions $W_0$, $b_0$ and
$\alpha_0$ characterizing this solution have the index $0$. We replace
the variable $\phi$, dilaton potential, by the running coupling
$\alpha_0$.

The system Eqs.~(\ref{eq:em01})-(\ref{eq:em03}) has been solved
numerically in Ref.~\cite{Galow:2009kw} by considering the bulk
coordinate at $z_* = 0.3426\,\textrm{GeV}^{-1}$ which is mapped to the
energy $E_* = 3\,\textrm{GeV}$ by the metric factor $b_0$ and the
arbitrary scale $\Lambda_E=0.264 \, \textrm{GeV}$. The choice of scale
size $\Lambda_E$ has only historical reasons, and in principle the mapping
of the energy coordinate to the z-coordinate has an arbitrary
constant.  The resulting parametrization of the $\beta$-function and
dilaton potential would then also change. We use this scale, because
we did not want to recalculate the fit to the running coupling and the
string tension with another energy scale. The value of the running
coupling at $E_* = 3\, \textrm{GeV}$ follows from the experimental PDG
data of the running coupling, c.f. Ref.~\cite{PDG}, and it is
$\alpha_0(z_*)= 0.25241$. The value of $W_0$ comes from the Einstein
equations which allow an explicit solution of the superpotential in
terms of $V(\alpha)$ and $\beta(\alpha)$,

\begin{eqnarray}
b_0(z_*)      &=& E_*/\Lambda_E \,, \label{eq:b0input} \\
\alpha_0(z_*) &=& 0.25241 \,, \label{eq:alpha0input} \\
W_0(z_*)      &=& \frac{9 \sqrt{- 3 V(\alpha_0(z_*))}} {8 \sqrt{9-\left(\frac{\beta(\alpha_0(z_*))}{\alpha_0(z_*)}\right)^2}}  \,. \label{eq:W0input}
\end{eqnarray}

We have solved the above equations and verified that they give the
same solutions as obtained in the previous paper on zero temperature
Ref.~\cite{Galow:2009kw}. Note the finite temperature $\beta = 1/T$ does not
enter in the gravity equations dependent on $z$.  The thermal gas is
solely defined by the periodicity in $\tau$. The initial conditions fix
the running of the coupling in the bulk and its scale $\Lambda$ 
which we compute from the numerical solution. Given the perturbative
$\beta$-function, which is parametrized for higher orders in Ref.~\cite{Galow:2009kw}
\begin{equation}
\beta(\alpha) = -\beta_0\alpha^2 - \beta_1\alpha^3- \beta_2\alpha^4  
+ \dots \,,
\end{equation}
the running coupling has the form:

\begin{eqnarray}
\alpha_0(z) &=& \frac{1}{\beta_0 L_z} +
\left(-\frac{\beta_1}{\beta_0^3}\log(L_z) + k\right)\frac{1}{L_z^2} +
\nonumber \\ &&+ \left(\frac{\beta_1^2}{\beta_0^5}(\log(L_z))^2 -
\frac{\beta_1}{\beta_0^5}(\beta_1+2\beta_0^3 k)\log(L_z) +
c^\alpha_3\right) \frac{1}{L_z^3}  + {\cal O}(L_z^{-4})\label{eq:luv} \,,
\end{eqnarray}
with
\begin{equation}
c^{\alpha}_3 = \frac{4}{9\beta_0}
+ \beta_0 k^2 + \frac{\beta_1}{\beta_0^2} k -
\frac{\beta_1^2}{\beta_0^5} +
\frac{\beta_2}{\beta_0^4}\,, \label{eq:ca3}
\end{equation}
and the definition
\begin{equation}
L_z :=  -\log(z\Lambda)\,.  \label{eq:defLz}
\end{equation}

The ultraviolet expansion of $b_0(z)$ follows from Eqs.~(\ref{eq:em02}) and (\ref{eq:em03}), using for $\alpha_0$ the expansion given by Eq.~(\ref{eq:luv}). Then $b_0$ reads
\begin{eqnarray}
b_0(z) &=& \frac{\ell}{z} \Bigg[  1 -\frac{4}{9} \beta_0 \alpha_0(z) +\frac{2}{81}\left(22\beta_0^2 -9\beta_1 \right) \alpha_0^2(z)  \nonumber \\
&&\qquad - \frac{4}{2187}\left(602\beta_0^3 - 540\beta_0\beta_1 + 81\beta_2 \right)\alpha_0^3(z)
+ {\cal O}(\alpha_0^4) \Bigg] \,. \label{eq:b0uv1}
\end{eqnarray}

In order to compute accurately the value of $\Lambda$, one can consider $b_0$ as a function of $\alpha_0(z)$ and expand it around the point $\alpha_0(z_*)=\alpha_*$ where the initial conditions were given. The expansion writes

\begin{eqnarray}
b_0(\alpha_0)&=& \frac{E_*}{\Lambda_E} \Bigg[ 1 +
\frac{1}{\beta(\alpha_*)} (\alpha_0-\alpha_*) +
\frac{(1-\beta^\prime(\alpha_*))}{2(\beta^\prime(\alpha_*))^2}
(\alpha_0-\alpha_*)^2 \nonumber \\
&+&\frac{(1-3\beta^\prime(\alpha_*) +
2(\beta^\prime(\alpha_*) -
\beta(\alpha_*)\beta^{\prime\prime}(\alpha_*))^2)}{6(\beta(\alpha_*))^2}
(\alpha_0-\alpha_*)^3 + \cdots\Bigg]\,. \label{eq:bz0as}
\end{eqnarray}
For $N_f=0$ a fit of the numerical solutions for $\alpha_0(z)$ and
$b_0(z)$ to the form given in Eqs.~(\ref{eq:luv}) and (\ref{eq:bz0as})
respectively, yields in both cases $\Lambda=0.543\,\textrm{GeV}$ which
is a factor two larger than $1/\ell = 0.228 \, \textrm{GeV}$. The
parameter $k$ is an integration constant which appears in the
Gell-Mann-Low integral~\cite{Vladimirov:1979my}. Note that a
particular choice of $k$ fixes the definition of $\Lambda$.  We have
chosen in this computation~$k=0$.

\section{Black hole solution of the Einstein Equations }
\label{sec:bhsol}

The gravitational equations have two different types of solutions for
the metric.  Besides the thermal gas solution which we discussed in
the previous section, there is also a solution which has a horizon
localized in the bulk coordinate at $z=z_h$ similar to the situation
in 4-dim gravity.  The phenomenology of the gluon plasma arises from
the competition of the free energies computed in both metrics. When
the free energy of the black hole solution has a lower value than the
thermal gas solution, the phase transition to the quark gluon plasma
takes place~\cite {Hawking:1982dh}.

Now we will discuss the equations of motion at finite temperature using the
black hole metric. The procedure is similar as discussed above for the
zero temperature case, but a new equation appears due to the black
hole factor $f(z)$. The black hole metric in Einstein frame has the form:

\begin{eqnarray}
G_{\mu\nu} &=& b^2(z) \cdot
\textrm{diag}\left(f(z),1,1,1,\frac{1}{f(z)}\right) \,,
\label{eq:Gbh}  \\
ds^2 &=& b^2(z)\left(f(z) d\tau^2 + dx_kdx^k +
\frac{dz^2}{f(z)}\right) \,, \label{eq:bhmetric}
\end{eqnarray}

where 
\begin{eqnarray}
f(0)&=1 \,, \\
f(z_h)&=0 \,.
\end{eqnarray}

Near the horizon the metric is given by

\begin{equation}
ds^2 = b^2(z_h)\left(f'(z_h)\cdot(z-z_h) d\tau^2 + d\vec{x} 
	\cdot d\vec{x}
	+ \frac{dz^2 }{f'(z_h)\cdot(z-z_h)}\right) \,.
\end{equation}

We define a new variable $\rho := \sqrt{z-z_h}$. In terms of $\rho$ we
obtain 

\begin{equation}
ds^2 = \frac{4 b^2(z_h)}{f'(z_h)}\left(\rho^2 \left( \frac{f'(z_h)}{2} d\tau \right)^2 + d\rho ^2 \right) + b^2(z_h)
d\vec{x} \cdot d\vec{x}\,.  
\end{equation}

The $\tau$-$\rho$ portion of the
metric defines a two-plane in polar coordinates with $\tau$ serving as
the angular coordinate.  To avoid a conical singularity at $\rho=0$ we
must require that 
$|f'(z_h)\tau/2|$ has a period of $2\pi$. In
Matsubara (imaginary-time) formalism the period is equal to inverse
temperature $\beta=T^{-1}$.  Thus, the temperature of a black hole
solution is given by \cite{Carlip:2008wv,PhysRevD.15.2752}

\begin{equation}
	T=-{f'(z_h) \over 4\pi } \,.
	\label{Eq:BHTemperature}
\end{equation}

The Einstein tensor 
$E_{\mu\nu}$ and energy momentum tensor $T_{\mu\nu}$ have different components in $00$, $44$ and spatial directions, 
and can be expressed in terms of $b(z)$, $f(z)$, dilaton field $\phi(z)$, dilaton potential $V(\phi)$ and its derivatives. One ends up with the four equations relevant at $T \ne 0$

\begin{eqnarray}
&&W' = \frac{16}{9} b W^2 - \frac{1}{f} \left( W f' - \frac{3}{4} b V \right) \,, \label{eq:em1} \\
&&b' = -\frac{4}{9} b^2 W \,, \label{eq:em2} \\
&&\alpha' = \alpha \sqrt{b W'} \,, \label{eq:em3} \\ 
&&f'' = \frac{4}{3} f' b W \label{eq:em4} \,,
\end{eqnarray}

where we have introduced the superpotential at finite temperature which is 
defined in analogy with Eq.~(\ref{eq:defW0}),

\begin{equation}
W(z) := -\frac{9}{4} \frac{b'(z)}{b^2(z)} \,. \label{eq:defW} 
\end{equation}

Note that the system of Eqs.~(\ref{eq:em1})-(\ref{eq:em4}) reduces to
the zero temperature formulas Eqs.~(\ref{eq:em01})-(\ref{eq:em03})
when $f\equiv 1$. Our prescription is to use the same dilaton
potential $V(\phi)$ at zero and finite temperature. Note, however,
that $V(z)$ is affected by the temperature dependence of
$\alpha(z)=e^{\phi(z)}$, i.e. $V(z) = V(\alpha(z))$.  In order to
solve the system of equations~(\ref{eq:em1})-(\ref{eq:em4}) one should
specify five initial conditions, as one has to handle three
differential equations of first order,
Eqs.~(\ref{eq:em1})-(\ref{eq:em3}), and one differential equation
Eq.~(\ref{eq:em4}) of second order.  The Eq.~(\ref{eq:em4}) for the
black hole function $f(z)$ can be solved analytically in terms of
$b$. Since our set up includes asymptotic freedom, $b(z)$ is different
from the conformal solution $b_{\textrm{\tiny conf}}(z)=\ell/z$. Also
the black hole factor deviates from the simple form of the conformal
solution $f_{\textrm{\tiny conf}}(z)=1-z^4/z_h^4$. Using
Eq.~(\ref{eq:em2}) one gets

\begin{equation}
f(z) = 1- C_f \int_0^z \frac{du}{b^3(u)} \,, \qquad C_f = \frac{1}{\int_0^{z_h}\frac{du}{b^3(u)}} \,, \label{eq:fzCf}
\end{equation}
where the first integration constant $C_f$ has been chosen such that
$f(z_h) = 0$.  The second integration constant is fixed to unity by
the requirement $f(0)=1$. From Eq.~(\ref{Eq:BHTemperature}) and the
definition of $C_f$ given by Eq.~(\ref{eq:fzCf}), one derives easily
the useful relation

\begin{equation}
C_f = 4\pi T b^3(z_h) \,. \label{eq:Cfb3}
\end{equation}
  
\begin{figure}[tbp]
\begin{center}
\epsfig{figure=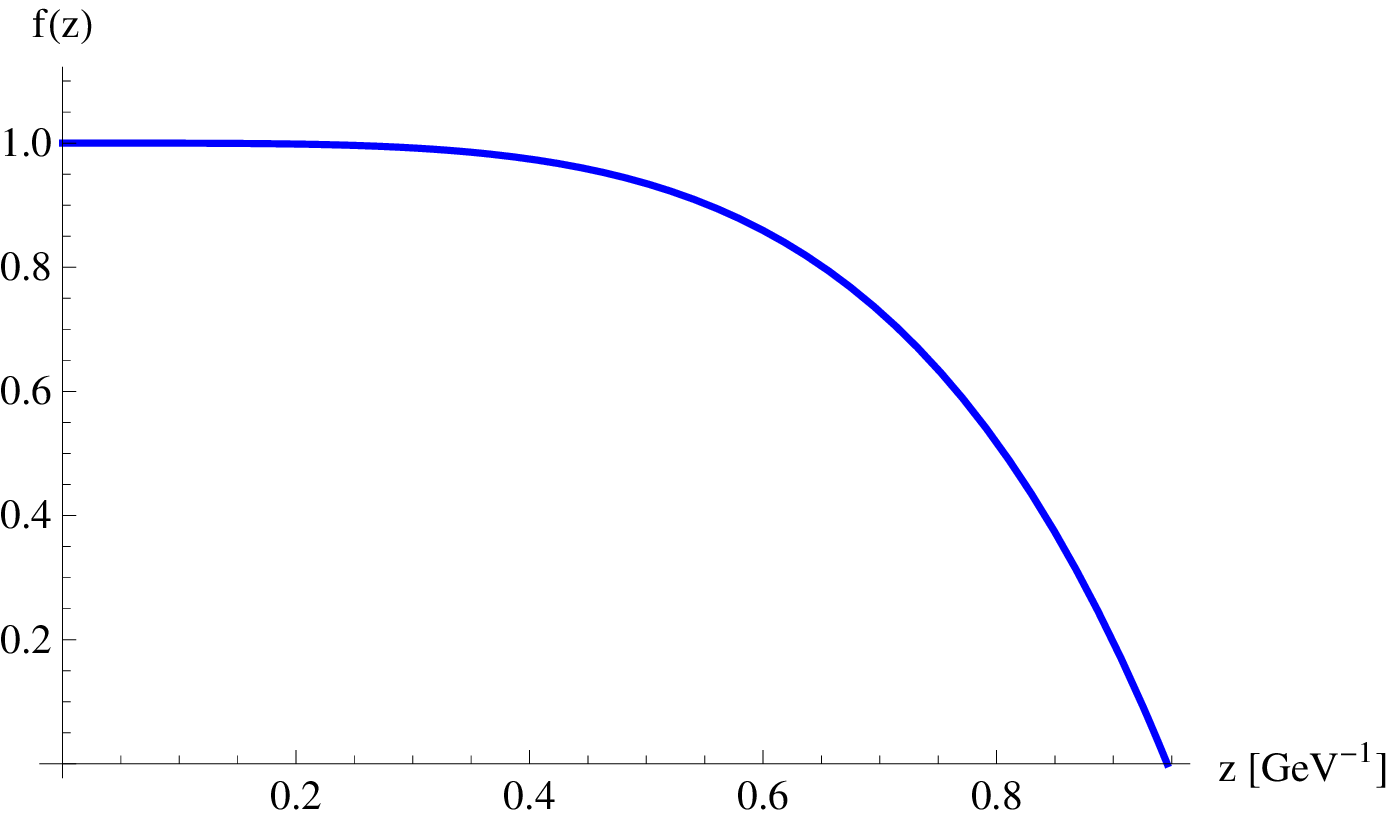,height=6.5cm,width=8.5cm}
\end{center}
\caption{$f$ as a function of $z$. We show the finite temperature result at $T=368\,$MeV.
}
\label{fig:fz}
\end{figure}

The technical procedure to solve numerically the Einstein equations is
explained in details in Appendix~A.  We show in Figs.~\ref{fig:Wz},
\ref{fig:bz}, \ref{fig:lambdaz} and \ref{fig:fz} the solutions for
$W(z)$, $z\cdot b(z)/\ell$, $\alpha(z)$ and $f(z)$ obtained at the temperature
$T= 368\,$MeV and compare them in the same figures with the zero
temperature solutions as computed in Sec.~\ref{sec:einsteinequations}. 
Note that the $T=0$ and $T \neq 0$ solutions
agree in the ultraviolet, i.e. $z < 0.5 \,\textrm{GeV}^{-1}$, but
differ in the infrared due to the thermal fluctuations introduced by the black hole horizon at $z_h=0.946\,\textrm{GeV}^{-1}$. 

\begin{figure}[tbp]
\begin{center}
\epsfig{figure=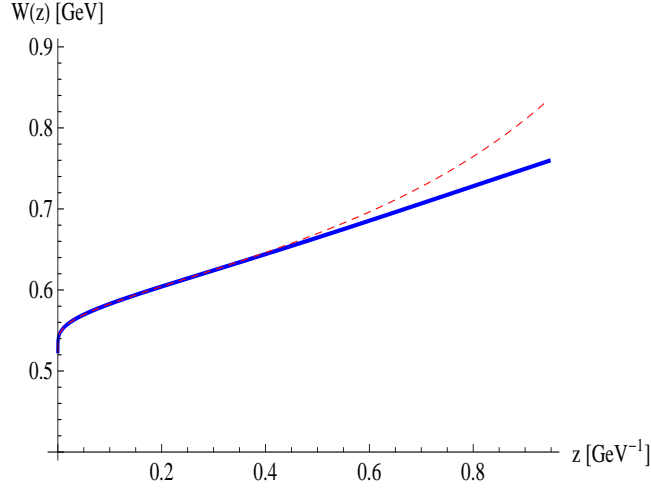,height=6.5cm,width=8.5cm}
\end{center}
\caption{Superpotential $W$ as a function of $z$. The full (blue) line gives the finite temperature result at $T=368 \,$MeV, and the dashed (red) line the zero temperature superpotential. The maximum value of $z$ shown in the plot corresponds to the horizon $z_h(T)=0.946\,\textrm{GeV}^{-1}$.
}
\label{fig:Wz}
\end{figure}

\begin{figure}[tbp]
\begin{center}
\epsfig{figure=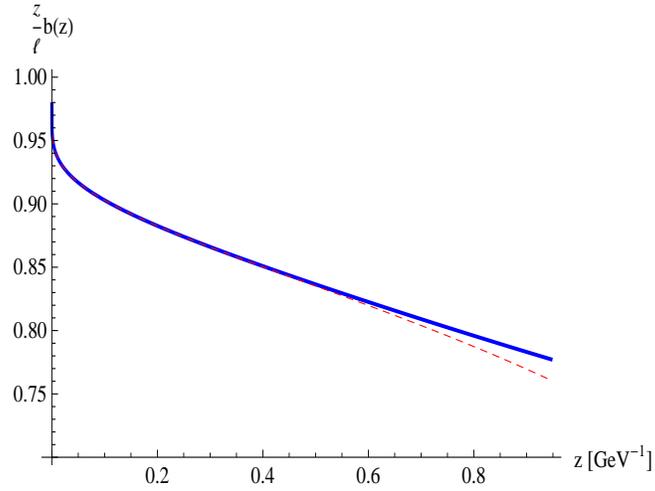,height=6.5cm,width=8.5cm}
\end{center}
\caption{Scale factor $b$ divided by the conformal limit $\ell/z$, as a function of $z$. We show the finite temperature result at $T=368\,$MeV and the zero temperature one. See Fig.~\ref{fig:Wz} for convention.
}
\label{fig:bz}
\end{figure}

\begin{figure}[tbp]
\begin{center}
\epsfig{figure=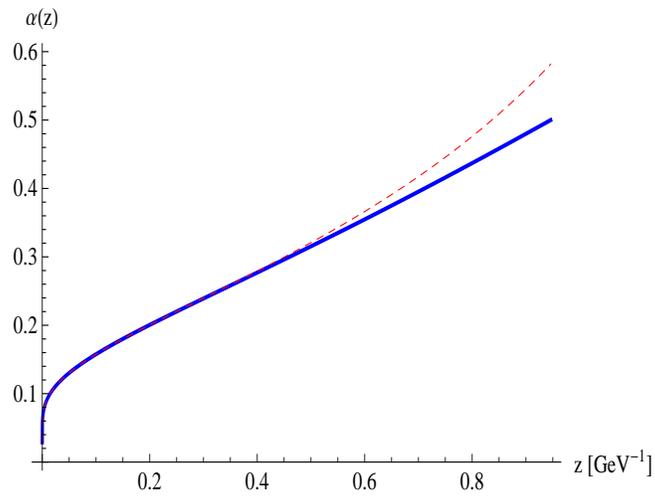,height=6.5cm,width=8.5cm}
\end{center}
\caption{Running coupling $\alpha$ as a function of $z$. We show the finite temperature result at $T=368\,$MeV and the zero temperature one. See Fig.~\ref{fig:Wz} for convention.
}
\label{fig:lambdaz}
\end{figure}

As we will explain in Sec.~\ref{sec:freeenergyEH}, we use $\alpha(z)$
to extract the trace anomaly. The finite temperature solution differs
from the zero temperature one at order ${\cal O}(z^4)$.  Care has to
be taken to keep track of leading logarithmic effects which are not
usually considered, c.f.~Ref.~\cite{Gursoy:2008za}, but are important
if one wants to calculate the gluon condensate $G$.  The difference
between zero and finite temperature solutions is mainly given by the
gluon condensate or the trace anomaly which equals $G$ up to
normalization factors,

\begin{equation}
G = \frac{\pi G_5}{15} \frac{\beta(\alpha)}{\alpha^2} \left( \langle
\textrm{Tr} F^2_{\mu\nu} \rangle_T -  \langle
\textrm{Tr} F^2_{\mu\nu} \rangle_0   \right) \,. \label{eq:Gnorm}
\end{equation}
The subscripts $T$ and $0$ stand for the thermal and vacuum values of
$\langle \textrm{Tr} F^2_{\mu\nu} \rangle$ respectively. The expressions relating zero and finite temperature quantities read up to ${\cal O}(z^4 \alpha_0^2)$

\begin{eqnarray}
b(z) &=& b_0(z) \left[ 1 + \frac{G}{\ell^3}z^4\left( 1 + c^b_1
\alpha_0(z) + c^b_2 \alpha^2_0(z) \right) + \cdots \right] \,,
\label{eq:bz4} \\ \alpha(z) &=& \alpha_0(z)\left[ 1 - \frac{45}{8}
\frac{G}{\ell^3\beta_0\alpha_0(z)}z^4 \left( 1 + \left(c^b_1 +
\frac{\beta_0}{4} - \frac{\beta_1}{\beta_0}\right)\alpha_0(z) +
c^\alpha_2 \alpha^2_0(z) \right) + \cdots \right]\,, \label{eq:az4} \\
W(z) &=& W_0(z) \left[ 1 - \frac{5G}{\ell^3} z^4 \left( 1 +
c^b_1\alpha_0(z) + \left( c^b_2 + \frac{\beta_0}{5} c^b_1 -
\frac{16}{45}\beta_0^2 \right)\alpha^2_0(z) \right) + \cdots \right]
\,, \label{eq:Wz4} \\ f(z) &=& 1 - \frac{z^4 Q_f(z)}{z_h^4 Q_f(z_h)}
\frac{1 - \frac{3}{2}\frac{G}{\ell^3} z^4 \left( 1 + c^b_1 \alpha_0(z)
+ \left( c^b_2 -\frac{\beta_0}{8}c^b_1 + \frac{\beta_0^2}{6}
\right)\alpha^2_0(z) \right) }{1 - \frac{3}{2}\frac{G}{\ell^3} z_h^4
\left( 1 + c^b_1 \alpha_h + \left( c^b_2 -\frac{\beta_0}{8}c^b_1 +
\frac{\beta_0^2}{6} \right)\alpha^2_h \right) } + \cdots \,,
\label{eq:fz4}
\end{eqnarray}
where
\begin{equation}
c^\alpha_2 = c^b_2 + c^b_1 \left( \frac{9}{20}\beta_0  - \frac{\beta_1}{\beta_0} \right) - \frac{4}{15}\beta_0^2 + \frac{\beta_1}{4} + \frac{\beta_1^2}{\beta_0^2} -\frac{\beta_2}{\beta_0} \,, \label{eq:ca2}
\end{equation}
being $Q_f(z)$ given by (c.f. Appendix B)
\begin{equation}
Q_f(z) = 1+\frac{4}{3}\beta_0 \alpha_0(z) - \frac{1}{9}(7\beta_0^2-6\beta_1)\alpha_0^2(z) + {\cal O}(\alpha_0^3) \,. \label{eq:Qf}
\end{equation}
 Higher orders in $\alpha_0$ and $z$ in Eqs.~(\ref{eq:bz4})-(\ref{eq:fz4}) are
indicated by dots.

To arrive at Eqs.~(\ref{eq:az4})-(\ref{eq:fz4}) we have assumed for
$b(z)$ an expansion of the form given by Eq.~(\ref{eq:bz4}), and used
the equations of motion Eqs.~(\ref{eq:em2})-(\ref{eq:em3}). Note that
on the r.h.s. of these expressions, inside the brackets, one can
substitute $\alpha_0(z)$ by $\alpha(z)$ and the expressions remain
valid at this order, as the difference between both quantities,
$\alpha_0(z)$ and $\alpha(z)$, is ${\cal O}(z^4)$,
c.f. Eq.~(\ref{eq:az4}). To get the values of the coefficients $c^b_1$
and $c^b_2$, one has to substitute the expansions for $b(z)$, $W(z)$
and $f(z)$ into the first equation of motion, Eq.~(\ref{eq:em1}), and
use the assumption that the dilaton potential at finite temperature as
a function of the dilaton field has the same functional form as the
one at zero temperature, i.e. $V_T(\alpha) =
V_{T=0}(\alpha_0)|_{\alpha_0=\alpha}$. Then one gets
\begin{eqnarray}
c^b_1 &=& \frac{19}{12}\beta_0 \,, \label{eq:cb1}  \\
c^b_2 &=& -\frac{263}{720}\beta_0^2 + \frac{7}{6}\beta_1 - \frac{C_f}{180 G} \beta_0^2 \,,  \label{eq:cb2}
\end{eqnarray}
where $C_f$ is defined in Eq.~(\ref{eq:fzCf}). Note that the
expression of $c^b_2$ in Eq.~(\ref{eq:cb2}) means that in the
expansion of the quantities in powers of $z$,
Eqs.~(\ref{eq:bz4})-(\ref{eq:fz4}), there are contributions not only of
the gluon condensate $G$, but also of $C_f$. The leading contribution
involving $C_f$ is $\sim z^4 C_f \alpha_0^2(z) $ times the
corresponding quantity at zero temperature ($\sim z^4 C_f \alpha_0(z)$
in the case of $\alpha(z)$).

The contribution of $G$ in Eqs.~(\ref{eq:bz4})-(\ref{eq:fz4}) is
visible in Figs.~\ref{fig:Wz}-\ref{fig:lambdaz}, and all these figures
consistently show that $G$ is positive.  Note that in our set-up the
correction arising from the NLO and NNLO coeficients of the $z^4$-term
are not small in the infrared, since $\alpha_0(z_h) \approx 0.5$ at
$T=368 \,\textrm{MeV}$.  Therefore we have to resort to a UV-analysis
to determine the gluon condensate G from the computation of
$\alpha(z)$.

\section{Free energy from the Einstein-Hilbert action}
\label{sec:freeenergyEH}

To get the thermodynamics of the five dimensional gravity-dilaton
model of Eq.~(\ref{eq:action5D}) one computes the free energy at fixed
temperature by introducing a lower cut off $z=\epsilon$ 
in the integral over the on shell action:

\begin{equation}
\beta {\cal F} = {\cal S}_{\textrm{\tiny reg}}(\epsilon)
\,. \label{eq:FSreg}
\end{equation}
Regularization is needed due to ultraviolet divergences near the
holographic boundary $\epsilon \to 0$.  The procedure to compute the regularized action with the black hole solution is explained in details in Ref.~\cite{Gursoy:2008za}. The free energy is computed as the difference between the free energy of the black hole solution and that of the thermal gas solution, so by definition the later has zero free energy. The result for the free energy is~\cite{Gursoy:2008za}

\begin{equation}
{\cal F} = \frac{1}{\beta} \lim_{\epsilon \to 0} ( {\cal
S}^{\textrm{\tiny BH}}_{\textrm{\tiny reg}}(\epsilon) - {\cal
S}^{\textrm{\tiny TG}}_{\textrm{\tiny reg}}(\epsilon) ) = \frac{\textrm{Vol}(3)}{16\pi G_5} \left( 15 G -
\frac{C_f}{4} \right) \,. \label{eq:Fk}
\end{equation}

\begin{figure}[tbp]
\begin{center}
\epsfig{figure=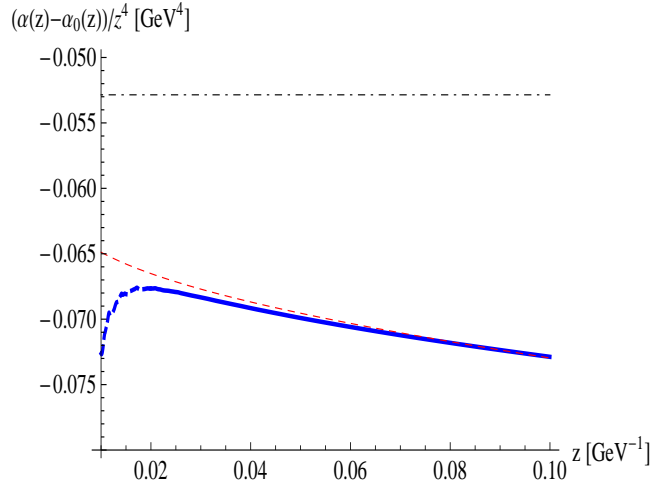,height=6.5cm,width=8.5cm}
\end{center}
\caption{Difference between the running coupling at finite and zero
temperature divided by $z^4$. We consider $T=368\,\textrm{MeV}$. Blue
line corresponds to the numerical computation from
Secs.~\ref{sec:einsteinequations} and \ref{sec:bhsol}. We plot as a
dashed blue line the regime $z<0.02\,\textrm{GeV}^{-1}$ which is
affected by numerical errors. Dashed red line corresponds to the
r.h.s. of Eq.~(\ref{eq:aa0dif})  including up to
${\cal O}(\alpha_0^2)$. We also show as a dashed
dotted black line the r.h.s. of Eq.~(\ref{eq:aa0dif}), but neglecting all the orders in $\alpha_0$, i.e. $-45G/(8\ell^3\beta_0)$.  }
\label{fig:plotaa2}
\end{figure}

From ${\cal F}$ one can compute the pressure and the rest of
thermodynamic quantities by applying the thermodynamic relations. The
values of $C_f$ and $G$ in the ultraviolet are given by (c.f. Appendix
B)

\begin{eqnarray}
C_f &=& \frac{4\ell^3}{z_h^4 Q_f(z_h)} = 4\pi^4 \ell^3 T^4 \Bigg[
1-\frac{4}{3}\beta_0 \alpha_h +
\frac{1}{9}(11\beta_0^2-6\beta_1)\alpha_h^2 + {\cal O}(\alpha_h^3)
\Bigg] \,, \label{eq:Cfuv} \\ G &=& \frac{\pi^4\ell^3}{45} T^4 \Bigg[
\beta_0^2 \alpha_h^2 + {\cal O}(\alpha_h^3) \Bigg] \,. \label{eq:Guv}
\end{eqnarray}
Inserting Eqs.~(\ref{eq:Cfuv}) and (\ref{eq:Guv}) into
Eq.~(\ref{eq:Fk}) one gets the UV assymptotic expansion of the free
energy which will be discussed later.  To deal with Eq.~(\ref{eq:Fk})
at temperatures near $T_c$ one has to compute the temperature
dependence of $C_f$ and $G$ numerically. Using Eq.~(\ref{eq:Cfb3}) one
can calculate $C_f$ from the numerical result of $b(z)$. On the other
hand the temperature dependence of $G$ can be computed by comparing
the thermal gas and black hole solutions in the ultraviolet, using
Eqs.~(\ref{eq:bz4})-(\ref{eq:fz4}). In the UV we perform a fit of the
difference $\alpha(z) - \alpha_0(z)$, and compute the coefficient $G$
for different temperatures. Note that it is much more efficient to use
$\alpha$ instead of $b$, as the latter diverges in the UV making it
more difficult to get reliable results for $G$. In this paper we have
analyzed carefully the expansion of the ${\cal O}(z^4)$ term in
$\alpha(z)-\alpha_0(z)$, Eq.~(\ref{eq:az4}), which gives the trace
anomaly $G$ in the plasma.  Higher order terms in $\alpha_0$ affect
appreciably the fit of $G$, and it is indispensable to consider at
least the order ${\cal O}(\alpha_0)$ to get good agreement of the
thermodynamic quantities with the numerical results from the
Bekenstein-Hawking entropy formula as computed in
Sec.~\ref{sec:Bekenstein_Hawking}. Up to ${\cal O}(\alpha_0^2)$ it
reads

\begin{equation}
\frac{\alpha(z) - \alpha_0(z)}{z^4} \simeq -\frac{45}{8} \frac{G}{\ell^3\beta_0}
\left( 1 + \left(c^b_1 + \frac{\beta_0}{4} -
\frac{\beta_1}{\beta_0}\right)\alpha_0(z) + c^\alpha_2 \alpha^2_0(z) \right) \,, \label{eq:aa0dif}
\end{equation}
where $c^b_1$ and $c^\alpha_2$ are given by Eqs.~(\ref{eq:cb1}) and
(\ref{eq:ca2}) respectively. In Fig.~\ref{fig:plotaa2} we show  the
numerical and analytical results of $(\alpha(z)-\alpha_0(z))/z^4$ for small z. 
Note that this quantity is not flat in this region, 
and a rough fit with a constant
term $\sim G$ (constant in $z$) leads in general to an overestimation of
the value of $G$, and then also on the value of ${\cal F}$,
c.f. Eq.~(\ref{eq:Fk}). This induces an appreciable error in the
behaviour of ${\cal F}$, and in the value of~$T_c$. For instance,
if we performed the fit by neglecting all higher orders in $\alpha_0$ on
the r.h.s. of Eq.~(\ref{eq:aa0dif}), we would get for the transition
temperature $T_c = 298.7\,\textrm{MeV}$, instead of the correct value
$T_c = 273.0\,\textrm{MeV}$ for zero flavours.

\section{Free Energy from the Bekenstein-Hawking Entropy}
\label{sec:Bekenstein_Hawking}

One of the postulates of the gauge/string duality is that the entropy
of the hot gauge theory equals the Bekenstein-Hawking entropy of its
gravity dual. This opens a quick way to check the results of the
previous section independently.  The Bekenstein-Hawking entropy is
proportional to the (3-dimensional) area of the black hole at the horizon $r=
r_h=\ell^2/z_h$ with the metric Eq.~(\ref{eq:Gbh}).  In non conformal AdS/QCD the entropy writes

\begin{equation}
S(T)= \frac{\textrm{Vol}(3)}{4 G_5} b^3(r_h(T)) \,,  \label{eq:entropy}
\end{equation}
where $G_5$ is the gravitational constant in 5-dim and $b$ is the
metric factor in Einstein frame. The prefactor $1/(4G_5)= \frac {2
  (N_c^2-1)}{4 \pi \ell^3} $ in conformal theory is much too large for
pure QCD, since ${\cal N}=4$ supergravity includes extra degrees of
freedom, gluinos and scalars, besides gluons.

In order to map out an equation of state, one needs the location of
the horizon $r_h=\ell^2/z_h$ as a function of temperature. One finds
for a temperature above some minimum value $T_{\textrm{\tiny min}}$ in
general a solution with a small $r_h$ (small black hole) and a
solution with a large $r_h$ horizon (large black hole).  Only the
large black hole is stable, because its free energy is a minimum. The
large black hole solutions are used to calculate the entropy.

The free energy due to black holes must be calculated by combining the
entropy $S_B$ of big black holes and the entropy $S_S$ of small black
holes.  The free energy of the big black hole can then be computed as~\cite{Gursoy:2008za}

\begin{eqnarray}
{\cal F}_B &=& {\cal F}_S(\infty)- \int_{\infty}^{T_{min}} S_S dT- \int_{T_{min}}^T S_B dT\,, \label{eq:p1}
\end{eqnarray} 
where the unstable free energy from small black holes vanishes in the limit $T \to \infty$, i.e. ${\cal F}_S(\infty)=0$.

The calculation of the free energy in terms of the entropy is much
easier than the full calculation shown in
section~\ref{sec:freeenergyEH}. The reason is that the extraction of
the gluon condensate $G$ is rather subtle and we had to develop the
procedure indicated in the previous section to get a reliable value
for $G$.  The perfect agreement of both methods gives a good guarantee
that the numerical solutions are reliable.

\begin{figure}[tbp]
\begin{center}
\epsfig{figure=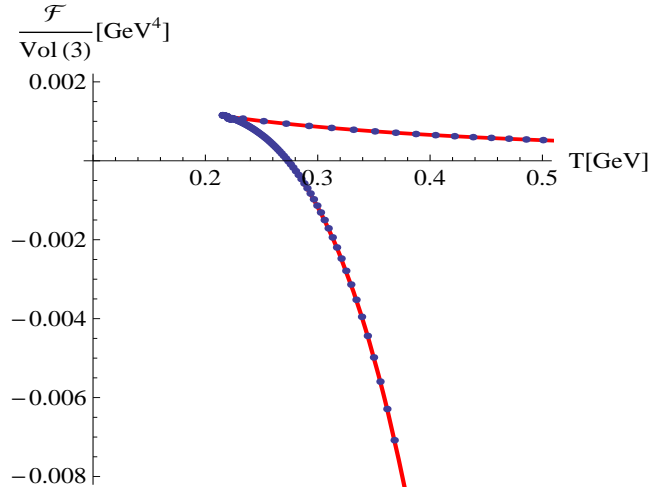,height=6.5cm,width=8.5cm}
\end{center}
\caption{Free energy density as a function of temperature.  We show as
  a full (red) line the result obtained starting from the
  Bekenstein-Hawking entropy formula,
  c.f. Sec.~\ref{sec:Bekenstein_Hawking}. Blue points correspond to
  the result using the Einstein-Hilbert action,
  c.f. Sec.~\ref{sec:freeenergyEH}, Eq.~(\ref{eq:Fk}). We include up to
  ${\cal O}(\alpha_0^2)$ in the r.h.s. of Eq.~(\ref{eq:aa0dif}) to
  compute $G$.  }
\label{fig:freeenergy12}
\end{figure}

We show in Fig.~\ref{fig:freeenergy12} the free energy obtained by
using the Bekenstein-Hawking entropy formula (full red line) as
computed in this section, and the Einstein-Hilbert action (blue
points) from Section~\ref{sec:freeenergyEH}, using NNLO terms in
$\alpha_0$ in Eq.~(\ref{eq:aa0dif}) to compute the gluon condensate
$G$ as a function of temperature. One clearly recognizes in this
figure the first order phase transition at the temperature $T_c=273
\,\textrm{MeV}$ for zero flavours, which is quite close to lattice
simulations. We consider the scale setting from the zero temperature
gravity as a great success for gauge gravity duality.  The upper
branch in Fig.~\ref{fig:freeenergy12} represents the small black holes
which are energetically disfavoured.

For high temperatures a weak coupling expansion of the pressure, $p =
-{\cal F}/\textrm{Vol}(3)$, can be made. We refer the reader to
Appendix~B for a detailed discussion of the ultraviolet properties of
the thermodynamic quantities.   For an analytical computation in the ultraviolet,
one makes  a weak coupling expansion in $\alpha_h=\alpha(z_h)$.
The coupling $\alpha_h$ evaluated at the black hole horizon is finite. One can relate $\alpha_h$ with the value of the running coupling at the scale $z=1/(\pi T)$ by the following equation:
 
\begin{equation}
\alpha_h = \alpha_T + \frac{\beta_0^3}{3}\alpha_T^4 + {\cal
O}(\alpha_T^5) \,, \qquad \alpha_T \equiv \alpha\left(z=\frac{1}{\pi
T}\right) \,, \label{eq:ahatex}
\end{equation}
where $\alpha_T$ is defined as indicated.  The easiest way to compute
the pressure $p(T)$ from the entropy density $s(T)$ is to solve the
equation $dp(T)/dT = s(T)$.  Then one can consider a general scheme
for the weak coupling expansion in the pressure, $p(T)/T^4 = p_0 +
p_1\alpha_h + p_2 \alpha_h^2 + p_3\alpha_h^3 + \cdots \,,$ and solve for
the coefficients in the above equation with $s(T)$ given by
Eq.~(\ref{eq:s1}).  $dp(T)/T$ can be computed easily by making use of
Eq.~(\ref{eq:dahdtex}). Then one can identify the coefficients in the
expansion. The result is

\begin{eqnarray}
\frac{p(T)}{T^4} &=& \frac{\pi^3 \ell^3 }{16 G_5} \Bigg[ 1 -
 \frac{4}{3} \beta_0 \alpha_h + \frac{2}{9} \left( 4\beta_0^2 -
 3\beta_1 \right) \alpha_h^2 -\frac{1}{162}\left( 91\beta_0^3
 -144\beta_0\beta_1 +72\beta_2 \right) \alpha_h^3 + {\cal
 O}(\alpha_h^4) \Bigg] \nonumber \\ &=& \frac{\pi^3 \ell^3}{16 G_5}
 \left[ 1 - 2.33 \alpha_h + 1.86 \alpha_h^2 + 0.33 \alpha_h^3 + {\cal
 O}(\alpha_h^4) \right] \,,
\label{eq:pwc}
\end{eqnarray} 
where we show in the last expression the values of the coefficients
corresponding to $N_c=3$ and $N_f=0$.

In AdS/QCD we may
choose the gravitational constant to reproduce the ideal gas limit at
high temperatures, i.e. $p(T)/T^4 \sim_{T\to \infty}
(N_c^2-1)\pi^2/45$. Then one gets

\begin{equation}
\frac{1}{16 G_5^{\infty}} = \frac{(N_c^2-1)}{45\pi\ell^3}\,. \label{eq:G5}
\end{equation}

We call the so determined constant $G_5^{\infty}$ indicating that it
follows from the large temperature limit. As pointed out in
Ref.~\cite{Alanen:2009xs}, this value is a factor $8/45$ smaller than
the value for ${\cal N}=4$ super Yang-Mills theory. This decrease may
be explained by the following two arguments. The number of degrees of
freedom is reduced in QCD compared with SQCD by a factor $(2/15)$.
QCD is weakly interacting at high energies compared with the AdS/CFT
theory which remains strongly interacting. This gives another factor
$(4/3)$.

\begin{figure}[tbp]
\begin{center}
\epsfig{figure=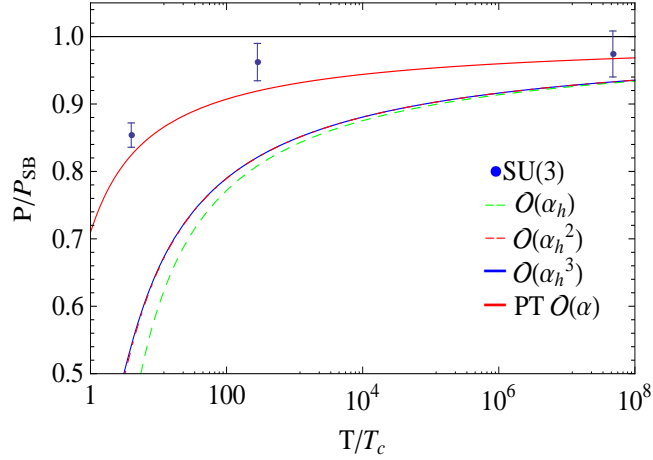,height=6.5cm,width=8.5cm}
\end{center}
\caption{Pressure normalized to the Stefan-Boltzmann limit, as a
function of $T/T_c$. We show as points the high temperature lattice
data for SU(3) taken from Ref.~\cite{Endrodi:2007tq}. We plot the
analytical result from the holographic model for several orders as an
ultraviolet expansion in powers of the running coupling, c.f. Eq.~(\ref{eq:pwc}), and also the QCD perturbative result up to order ${\cal O}(\alpha)$, c.f. Eq.~(\ref{eq:pqcdPT}). }
\label{fig:pressureFodor}
\end{figure} 

We show in Fig.~\ref{fig:pressureFodor} the pressure as a function of
temperature, normalized to the Stefan-Boltzmann limit. It is noteworthy and visible in the figures that the expansion in terms of $\alpha_h$ converge quite rapidly. On the other hand, as one can see the holographic model approaches the ultraviolet limit
slower than lattice data. The asymptotic expansion of the pressure in
QCD with $N_c=3,N_f=0$ has the form:

\begin{equation}
\frac{p_{\textrm \tiny QCD}(T)}{T^4} = \frac{8 \pi^2 }{45} \Bigg[ 1 - \frac{15}{4 \pi }\alpha
+30  \left(\frac{\alpha}{\pi}\right)^{3/2}+ \dots \Bigg]\,. \label{eq:pqcdPT}
\end{equation}

If one compares the  ${\cal O}(\alpha)$ coefficient in the
ultraviolet expansion of
the holographic model  ($p_1^{\textrm{\tiny AdS}}=\frac{44}{6 \pi}$) 
Eq.~(\ref{eq:pwc}) with  the corresponding coefficient
from perturbative QCD ($p_1^{\textrm{\tiny pQCD}} =\frac{15}{4 \pi}$),
one gets a factor $p_1^{\textrm{\tiny AdS}}/p_1^{\textrm{\tiny pQCD}}=
88/45 \simeq 1.956$. This ratio explains the deviation observed in
Fig.~\ref{fig:pressureFodor}, since the leading coefficient gives a good approximation
to lattice QCD at high temperatures
$T=(100-1000)T_c$.  The same factor appears when one
compares the first non-zero coefficient in the ultraviolet expansion
of the trace anomaly which is ${\cal O}(\alpha^2)$, and so one expects
that the holographic model also predicts values of
$(\epsilon-3P)/T^4$ larger than lattice data at high
temperatures.  This seems to be a property of the present
holographic model, and there is no easy way to cure it.  In
Ref.~\cite{Alanen:2009xs} a behaviour $\beta(\alpha)\sim -\alpha^q$
for $q>2$ has been studied and numerical consistency between QCD and
this AdS-model with a very simplified $\beta$ function is reached for
$q=10/3$. This parametrization disagrees, however, with the standard
perturbative behavior of the $\beta$-function of QCD, which has been
the basic starting point to make AdS similar to QCD in the program of
Refs.~\cite{Gursoy:2007cb,Gursoy:2007er,Gursoy:2008za}.

The parameterization presented in Ref.~\cite{Gursoy:2009jd} is
successful to reproduce lattice data in the regime $1 <T/T_c < 5$, but
in view of our analysis it is clear that this is because non
perturbative effects are much stronger than perturbative ones even at
very high temperatures, which seems to be not reasonable. As a matter
of fact, the $\beta$-function of Ref.~\cite{Gursoy:2009jd} agrees only
for $\alpha_s < 10^{-7}$ with an $1\%$ accuracy with the asymptotic
$\beta$-function of QCD. The dilaton potential used in this reference
has the correct uv-behaviour of QCD, but for all practical purposes it
is a model potential which is designed to fit the thermal equation of
the gluon plasma and not the $\beta$-function established in
perturbative QCD for $\alpha_s<0.2$. Since in Ref.~\cite{Gursoy:2009jd} the scheme dependent $\beta_2 \alpha^4$ term in the $\beta$-function is very large, it is simply not clear how to determine the running of $\alpha_s$ in the MOM or ${\overline {\textrm MS}}$ scheme from the parametrization of $\beta(\alpha)$ in this reference.

It is interesting to analyze the dependence of the phase transition
temperature when the number of flavors is changed. The authors of
Ref.~\cite{Braun:2009ns} have studied the $N_f$-dependence of the
transition temperature $T_c$ with the help of renormalization group
flow equations. As shown in this reference, the scale
$\Lambda_{\textrm{\tiny QCD}}$ changes with $N_f$. Therefore it is
recommended to solve the Einstein equations by keeping the running
coupling fixed at a mid scale $3\, \textrm{GeV}$,
c.f. Eq.~(\ref{eq:alpha0input}).  In order to study the flavour
dependence of $T_c$ in the present model, we vary the number of
flavors $N_f$ from $N_f=0$ to $N_f=10$ in the coefficients $\beta_0$
and $\beta_1$ of the $\beta$-function,
c.f. Eqs.~(\ref{eq:b0})-(\ref{eq:b1}), which control the short distance
(high temperature) regime.  We retain the nonperturbative parameters
$b_2$ and $\bar\alpha$ as in Eq.~(\ref{eq:b2alpha}), which is
reasonable because $b_2$ and $\bar\alpha$ are responsible for the
infrared large distance (low temperatue) regime, and the string
tension in the $\bar{q}q$ potential mainly depends on these two
parameters. We don't study the effect of dynamical quarks, therefore
our analysis is restricted to a quenched approach.

\begin{figure}[tbp]
\begin{center}
\epsfig{figure=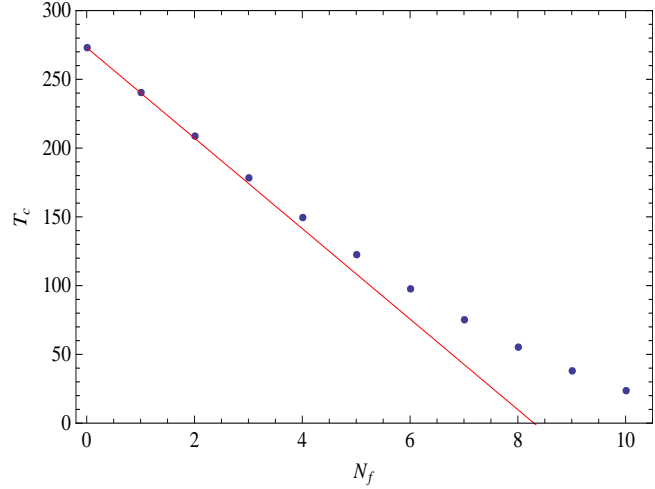,height=6.5cm,width=8.5cm}
\end{center}
\caption{Phase transition temperature as a function of the number of
flavors.  The points correspond to the numerical computation of the holographic model as explained in Secs.~\ref{sec:freeenergyEH} and~\ref{sec:Bekenstein_Hawking}, within a quenched approximation. We also plot as (red) continuous line the scaling law of Eq.~(\ref{eq:Tclinear}).}
\label{fig:Tcnf}
\end{figure}

Fig.~\ref{fig:Tcnf} shows the dependence of the critical temperature
$T_c$ on the number of flavors $N_f \le 10$. Larger values
of $N_f$ are difficult to implement numerically. We get for 
$N_f=0$ as  transition temperature
\begin{equation}
T_{N_f=0} = 273.0 \,\textrm{MeV} \,,\label{eq:T0e}
\end{equation}
 which is very close to the lattice results $T_c = 270(2) \,
\textrm{MeV}$~\cite{Beinlich:1997ia}. It is gratifying that the
absolute value of the transition temperature comes out so well inspite
of the slow convergence of the pressure towards the Stefan-Boltzmann limit
with increasing temperature.
Comparing results with different flavour numbers we obtain an almost
linear behaviour of the transition temperature for small $N_f$:
\begin{equation}
T_c = T_{N_f=0} \left( 1 - \kappa N_f + {\cal O}(N_f^2) \right)\,, \qquad \kappa = 0.1205 \,. \label{eq:Tclinear}
\end{equation}
This linear scaling of the critical temperature with $N_f$ for small
$N_f$ has been claimed in Refs.~\cite{Braun:2009ns,Braun:2009si}. The
value of $\kappa$ we get is quite close to the one estimated in
Ref.~\cite{Braun:2009ns}, $\kappa \simeq 0.107$.  The flattening of
the function $T_c(N_f)$ for high values of $N_f$ in
Fig.~\ref{fig:Tcnf} is in accordance with
Ref.~\cite{Braun:2009ns}. This reference explains this fact as a
consequence of the IR fixed-point structure of the theory. In our case
the $\beta$-function given by Eq.~(\ref{eq:betaparam}) does not have
an IR fixed-point, and the flattening is a consequence of the
weakening of the infrared coupling $\alpha$ when $N_f$ increases. This
means that it takes smaller scales to reach a critical coupling
$\alpha(T_c)$ to bind the gluons into glueballs, and therefore $T_c$
has to decrease. We found that for the points plotted in
Fig.~\ref{fig:Tcnf}, the value of $\alpha(T_c)$ is not affected much by $N_f$.

\section{Thermodynamic Observables independent of $G_5$}
\label{sec:thermodynamics}

We study in this section several thermodynamic quantities which are
independent of the 5-d gravitational constant $G_5$. 
For a more complete discussion of those thermodynamic quantities which
dependent on $G_5$  we refer to 
Ref.~\cite{Megias:2010tj}.

First we focus on the speed of sound~$c_s$. From the specific heat per unit volume $c_v$:
\begin{eqnarray}
c_v &=& T \frac{\partial^2 p}{\partial T^2}  \,,
\end{eqnarray}
and the entropy density $s$, one obtains the speed of sound:

\begin{eqnarray}
c_s^2 &=& \frac{s}{c_v}  \,.
\end{eqnarray}

A computation of this quantity in the ultraviolet leads to (see Appendix B for details)
\begin{eqnarray}
c_s^2 &=& \frac{s}{c_v} = \frac{1}{3} \Bigg[ 
 1 - \frac{4}{9}\beta_0^2 \alpha_h^2 + \frac{2}{9}\beta_0\left( \beta_0^2-4\beta_1\right)\alpha_h^3 + {\cal O}(\alpha_h^4)
\Bigg] \,. \label{eq:cstext}
\end{eqnarray}

\begin{figure}[tbp]
\begin{center}
\epsfig{figure=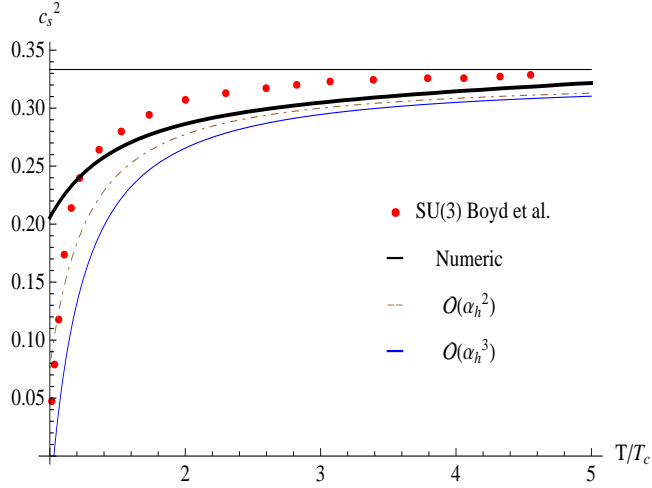,height=6.5cm,width=8.5cm}
\end{center}
\caption{Speed of sound squared as a function of $T/T_c$. We show as
points the lattice data for SU(3) taken from
Ref.~\cite{Boyd:1996bx}. The colored curves represent the analytical
result from the holographic model for several orders as an ultraviolet
expansion in powers of the running coupling, and the black solid line
refers to the full numerical result.  }
\label{fig:cs2}
\end{figure}

In Fig.~\ref{fig:cs2} we show the speed of sound computed with the holographic
model, and compare the result with lattice data of Ref.~\cite{Boyd:1996bx}. We
also plot  the analytical ultraviolet approximation given in
Eq.~(\ref{eq:cstext}) including several orders in an expansion in
$\alpha_h$. Since $c_s^2$ becomes close to $1/3$ in the calculation,
we see that we have massless excitations in the plasma in the
temperature range $2T_c<T< 5T_c$.

The spatial string tension is another quantity which is very useful to
test AdS/QCD models. It is non-vanishing even in the deconfined phase,
and it gives useful information about the non perturbative features of
high temperature QCD.  With a quark and an antiquark located at
$x=\frac{d}{2}$ and $x=-\frac{d}{2}$ respectively, the computation of
the correlation function of rectangular Wilson loops in the $(x,y)$
plane leads to a potential between quark and antiquarks which behaves
linearly at large distances, i.e.

\begin{equation}
\langle W[{\cal C}] \rangle \stackrel{y\to\infty} \simeq  e^{-y \cdot V(d)} \,, \qquad
V(d) \stackrel{d\to\infty} \simeq \sigma_s \cdot d  \,.
\end{equation}
For details on the computation we refer the reader to e.g. Ref.~\cite{Alanen:2009ej} and references
therein. The spatial string tension takes the following form:

\begin{equation}
\sigma_s(T) = \frac{1}{2\pi l_s^2} \alpha_h^{4/3} b^2(z_h) \,, \label{eq:sigmas1}
\end{equation}
where $l_s$ is the string length. Making use of all the technology
developed in Appendix~B, we can easily compute the UV asymptotics of
$\sigma_s(T)$. Using the ultraviolet expansion of $b(\alpha)$ given by
Eq.~(\ref{eq:bzuv}) and the corresponding expansion of $z_h$ given by
Eq.~(\ref{eq:Tzhuv}), then Eq.~(\ref{eq:sigmas1}) leads to

\begin{eqnarray}
\sigma_s(T) &=& \frac{\ell^2}{2 l_s^2} \pi T^2 \alpha_h^\frac{4}{3}
\Bigg[ 1 - \frac{8}{9}\beta_0\alpha_h
+\frac{2}{81}(25\beta_0^2-2\beta_1)\alpha_h^2 \nonumber \\
&&\qquad\qquad\qquad -\frac{1}{2187}\left( 931\beta_0^3
-1836\beta_0\beta_1 + 642\beta_2 \right)\alpha_h^3 + {\cal
O}(\alpha_h^4) \Bigg] \,. \label{eq:sigmaswc}
\end{eqnarray}

\begin{figure}[tbp]
\begin{center}
\epsfig{figure=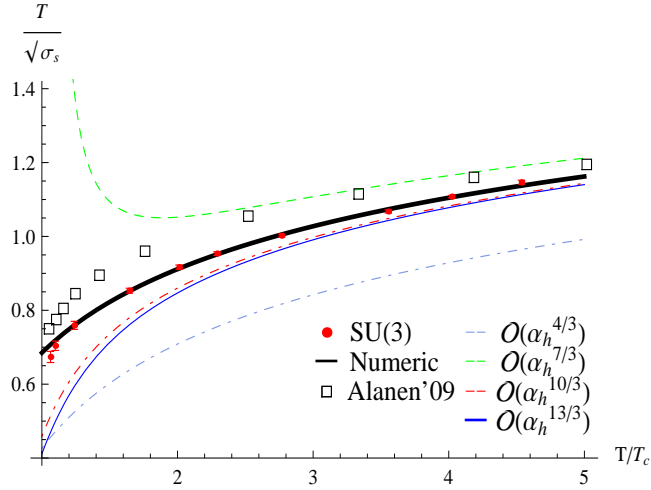,height=6.5cm,width=8.5cm}
\end{center}
\caption{$T/\sqrt{\sigma_s}$ as a function of temperature (in units of
  $T_c$). We show as filled (red) points the lattice data for SU(3)
  taken from Ref.~\cite{Boyd:1996bx}. The colored curves represent the
  analytical result from the holographic model for several orders as
  an ultraviolet expansion in powers of the running coupling,
  c.f. Eq.~(\ref{eq:sigmaswc}), and the black solid line refers to the
  full numerical result of Eq.~(\ref{eq:sigmas1}). We use the value
  $l_s = 1.94\,\textrm{GeV}^{-1}$. We show for comparison as square
  points the numerical result obtained from the model of
  Ref.~\cite{Alanen:2009ej}.  }
\label{fig:sigmas}
\end{figure}

We show in Fig.~\ref{fig:sigmas} a plot of $T/\sqrt{\sigma_s}$ as a
function of temperature including several orders in
Eq.~(\ref{eq:sigmaswc}), and the full numerical computation from
Eq.~(\ref{eq:sigmas1}). We show for comparison also the numerical
result obtained from the model of Ref.~\cite{Alanen:2009ej}. One can
see that our model reproduces very well the lattice data in the regime
$ 1.10 <T/T_c < 4.5$. A fit to the lattice data taken from
Ref.~\cite{Boyd:1996bx} gives a good $\chi^2/\textrm{d.o.f.} < 1$, and
it is obtained by using $l_s =1.94 \,\textrm{GeV}^{-1}$ which is
$30\%$ larger than the value quoted in Ref.~\cite{Galow:2009kw} based
on a joint analysis of the heavy $Q\bar{Q}$ potential and running
coupling at zero temperature. From a computation of the string tension
at zero temperature one can see that this increase in $l_s$ can be
partially explained as an effect of the change in the number of
flavors, as Ref.~\cite{Galow:2009kw} considers $N_f=4$ while we
consider here $N_f=0$. The string tension at $T=0$ can be computed
as~\cite{Alanen:2009ej}

\begin{equation}
\sigma = \frac{1}{2\pi l_s^2} b_0^2(z_*) \alpha_0^{4/3}(z_*) \,,
\label{eq:sigmaT0}
\end{equation}
where $z_*$ gives the minimum of  $b_0^2(z)\alpha^{4/3}(z)$. Using
the solution of $b_0$ and $\alpha_0$ for $N_f=0$, one reproduces the
physical value $\sigma \simeq (0.420 \,\textrm{GeV})^2$ for
$l_s^{N_f=0} = 2.22 \,\textrm{GeV}^{-1}$, while for $N_f=4$ one gets
$l_s^{N_f=4} = 1.45 \,\textrm{GeV}^{-1}$~\cite{Galow:2009kw}. The
discrepancy of $l_s^{N_f=0}$ with the value we get from the fit of
$\sigma_s(T)$ is $12\%$.

The vacuum expectation value of the Polyakov loop serves as an order
parameter for the deconfinement transition in gluodynamics. The
correlation function of two Polyakov loops taken in the large distance
limit leads to the vacuum expectation value of one single Polyakov
loop squared. This means that the Polyakov loop is related to the free
energy of a single quark~$F_q$ as
\begin{equation}
\langle {\cal P}(\vec{x}) \rangle = e^{-\beta F_q(\vec{x})} \,.
\end{equation}

 One of the main problems of the computation of this quantity is the renormalization. The multiplicative renormalizability of the Polyakov loop was first established in Ref.~\cite{Polyakov:1980ca}. The Polyakov loop was computed in perturbation theory for the first time in Ref.~\cite{Gava:1981qd} in pure gluodynamics. There has been recent progress to renormalize it in the lattice following different methods based on the computation of the one point and two point correlation functions of Polyakov loops. The multiplicative renormalization is then reached by identifying and extracting the quark self energy, see e.g. Refs.~\cite{Kaczmarek:2002mc,Gupta:2007ax}.  In Ref.~\cite{Megias:2005ve} it was proposed by one of the authors a phenomenological ansatz based on a dimension two gluon condensate of the dimensionally reduced effective theory of QCD, which was quite successful to reproduce lattice data of the Polyakov loop in the deconfined phase down to $T = 1.03 T_c$.~\footnote{See also Refs.~\cite{Megias:2007pq,Megias:2009mp} for an application of this ansatz to compute the heavy quark-antiquark free energy and the equation of state of QCD.} This ansatz follows from the observation for the first time in Ref.~\cite{Megias:2005ve} that close and above $T_c$ the behavior of the Polyakov loop is characterized by power corrections in $1/T^2$. These power corrections were later observed also in the equation of state of gluodynamics~\cite{Pisarski:2006yk}. The computation of the Polyakov loop within the AdS/QCD formalism was addressed recently in Ref.~\cite{Andreev:2009zk} within a model based on a specific choice of the warp factor $b(z)$ which naturally introduces these power corrections. In the following we will consider this approach, but using our model dictated by the 5-d grativy action.

One can compute the vacuum expectation value of the Polyakov loop from
the Nambu-Goto action of a string hanging down from a static quark on
the boundary into the bulk.~\footnote{The coupling of the two
  dimensional curvature $R^2$ to the dilaton field is a well known
  $\alpha^\prime$ correction to the Polyakov action which propagates
  to the Nambu Goto action~\cite{Kiritsis:2007zz}. In first order it
  enters there by a modified non conformal metric, as it is considered
  here. Higher order terms are interesting to investigate, but are in
  the scope of a separate longer study.} The fundamental string is
stretched between the test quark at the boundary ($z = 0$) and the
horizon ($z = z_h$) of the black hole solution. See
Ref.~\cite{Andreev:2009zk} for details. The Nambu-Goto action then
reads

\begin{equation}
S_{\textrm{\tiny NG}} = \frac{1}{2\pi l_s^2 T} \int_0^{z_h} dz \;
\alpha^{4/3}(z) b^2(z) \,. \label{eq:NG_action2}
\end{equation}

The action Eq.~(\ref{eq:NG_action2}) is divergent at $z=0$. One can
regularize it by substracting the action of the thermal gas solution
up to a cutoff $z_c$:

\begin{eqnarray}
S_{\textrm{\tiny NG}}^{\textrm{\tiny reg}} &=& \frac{1}{2\pi l_s^2 T}
\left[ \int_0^{z_h} dz \; \alpha^{4/3}(z) b^2(z) - \int_0^{z_c} dz \;
\alpha_0^{4/3}(z) b_0^2(z) \right] \nonumber \\ &=& \frac{1}{2\pi
l_s^2 T} \left[ \int_0^{z_h} dz \; \left( \alpha^{4/3}(z) b^2(z) -
\alpha_0^{4/3}(z) b^2_0(z) \right) - \int_{z_h}^{z_c} dz \;
\alpha_0^{4/3}(z) b_0^2(z) \right] \,.  \label{eq:Sngreg1}
\end{eqnarray}

The cutoff becomes necessary because the free energy of a single quark
diverges at $T=0$. Note that $z_c$ introduces a normalization constant
into the free energy $F_q = T \cdot S_{\textrm{\tiny
NG}}^{\textrm{\tiny reg}}$. In the second equality of
Eq.~(\ref{eq:Sngreg1}) we have divided the action corresponding to the thermal gas solution
into two integrals. The first integral inside
the bracket in Eq.~(\ref{eq:Sngreg1}) is UV convergent, as zero and
finite temperature solutions have the same behavior in the UV.  The
renormalized vacuum expectation value of the Polyakov loop then writes

\begin{equation}
L_R(T) = e^{-S_{\textrm{\tiny NG}}^{\textrm{\tiny reg}} }  \,. \label{eq:Lreg1}
\end{equation}

\begin{figure}[tbp]
\begin{center}
\epsfig{figure=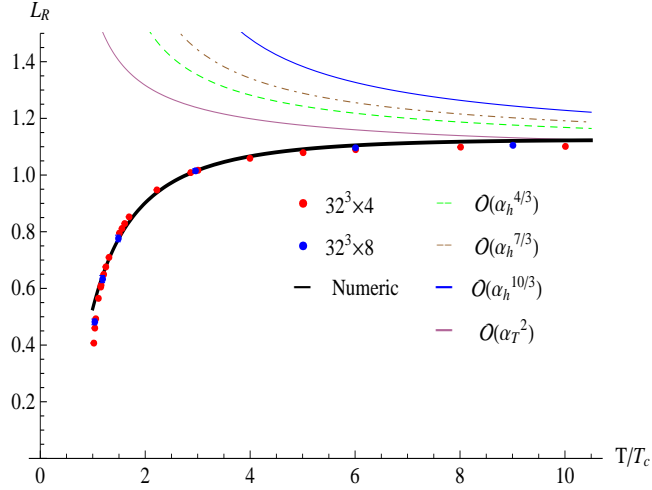,height=6.5cm,width=8.5cm}
\end{center}
\caption{Expectation value of the Polyakov loop as a function of temperature (in units of $T_c$). Full (black) line corresponds to the numerical computation of Eqs.~(\ref{eq:Sngreg1})-(\ref{eq:Lreg1}). We show as points lattice data for SU(3) taken from Ref.~\cite{Gupta:2007ax} for $N_\sigma^3\times N_\tau = 32^2\times 4$ and $32^3\times 8$. The colored curves represent the analytical result from the holographic model including several orders in $\alpha_h$, c.f. Eq.~(\ref{eq:L3uvtext}). We also plot the result from standard perturbative QCD up to ${\cal O}(\alpha_T^2)$ given by Eq.~(\ref{eq:Lqcduv}). We use the values $l_s = 2.36 \, \textrm{GeV}^{-1}$ and $z_c = 0.43\,\textrm{GeV}^{-1}$.}
\label{fig:PLnf3new2}
\end{figure}

$S_{\textrm{\tiny NG}}^{\textrm{\tiny reg}}$ defined in
Eq.~(\ref{eq:Sngreg1}) tends to zero in the limit $T \to \infty$
independently of the value of $z_c$, and so $L_R(T)$ tends to $1$.  We
show in Figure~\ref{fig:PLnf3new2} as a continuous black line the
behavior of $L_R$ as a function of temperature computed numerically
from Eqs.~(\ref{eq:Sngreg1})-(\ref{eq:Lreg1}), and its comparison with
lattice data for gluodynamics with $N_c=3$ taken from
Ref.~\cite{Gupta:2007ax}. In order to reproduce lattice data, we have
performed a fit by considering the string length $l_s$ and the cutoff
$z_c$ as free parameters. The best fit in the regime $T_c< T < 10 T_c$
leads to
\begin{equation}
l_s = 2.36 \, \textrm{GeV}^{-1}\,, \qquad  z_c = 0.43 \, \textrm{GeV}^{-1} \,. \label{eq:lszc}
\end{equation}
Note that $l_s$ is $20 \%$ larger than the value we used for the spatial string tension, but it differs just $5 \%$ from the value one needs to reproduce the string tension at $T=0$ (see Eq.~(\ref{eq:sigmaT0}) and discussion below).  Our
approach fits the Polyakov loop very well without a dimension two condensate,
since a dimension two operator would show up in the ultraviolet expansion of
the thermal solutions near $z=0$, Eqs.~(\ref{eq:bz4})-(\ref{eq:fz4}). This does not exclude that
a good fit to the data exists of the form $-2\log L_R \simeq a + b
(T_c/T)^2 $ with $a=-0.23$, $b=1.60$, in accordance with
Ref.~\cite{Megias:2005ve} (see also Ref.~\cite{Andreev:2009zk}). The Polyakov loop is zero in the confined phase and our approach gives a nonzero value at $T_c$ given by $L_R (T_c) = e^{-\frac{1}{2}(a+b)} = 0.50$. This first order jump is similar to the one predicted by the more reliable lattice data  $N_\sigma^3\times N_\tau = 32^3\times 8$, c.f.~Fig.~\ref{fig:PLnf3new2}.

We compute in details in Appendix~B the UV asymptotics of the Polyakov loop. The results is

\begin{equation}
L_R(T) = \exp\left[\frac{\ell^2}{2 l_s^2} \alpha_h^{\frac{4}{3}}
\left( 1 + \frac{4}{9}\beta_0\alpha_h +
\frac{1}{81}\left(161\beta_0^2+72\beta_1 \right)\alpha_h^2 + {\cal
O}(\alpha_h^3) \right)\right] \,. \label{eq:L3uvtext}
\end{equation}

We show in Fig.~\ref{fig:PLnf3new2} the analytical result given by
Eq.~(\ref{eq:L3uvtext}), using the value of $l_s$ quoted in
Eq.~(\ref{eq:lszc}). The Polyakov loop was computed in perturbative
QCD up to NLO in Ref.~\cite{Gava:1981qd} (see also Ref.~\cite{Megias:2005ve}), and it has been recently corrected by two groups Ref.~\cite{Burnier:2009bk,Brambilla:2010xn}. For gluodynamics this gives

\begin{equation}
L_{\textrm{\tiny PT}}(T) = \exp\left[ \frac{N_c^2-1}{2 N_c}
\sqrt{\pi}\, \alpha_T^\frac{3}{2} + \frac{N_c^2-1}{4} \left(
\log\alpha_T + \log(4\pi) + \frac{1}{2} \right) \alpha_T^2 + {\cal
O}(\alpha_T^{5/2}) \right] \,. \label{eq:Lqcduv}
\end{equation} 

Note that since the perturbative $\beta$-function starts at order
$\alpha^2$, changes in the scale $\mu$ affect ${\cal
O}(\alpha^{5/2})$. In Eq.~(\ref{eq:L3uvtext}) the power counting in
$\alpha_h$ doesn't follow the perturbative scheme. This discrepancy
with PT is common of all the renormalization group revised models
constructed by the general procedure of Kiritsis et al.,
c.f. Refs.~\cite{Gursoy:2007er,Gursoy:2007cb}. We have plotted in
Fig.~\ref{fig:PLnf3new2} also the perturbative result given by
Eq.~(\ref{eq:Lqcduv}). Note that lattice data approach the perturbative result
very accurately for $T$ above $8T_c$. Here the AdS-perturbation
theory does not seem to converge rather rapidly.

\section{Discussion and final remark}
\label{sec:discussion}

We have demonstrated in this work the numerical agreement between
computations of the free energy from the Einstein-Hilbert action and
from the Bekenstein-Hawking entropy formula. Both approaches leads to
the same result, but the former method is much more sensitive to
numerical errors, and an accurate computation is only possible when
one takes care of including leading logarithmic effects in an
ultraviolet expansion of the scale factor at finite temperature.

We have also computed analytical expressions in the ultraviolet for the
thermodynamic quantities as an expansion in powers of the running
coupling $\alpha_h$ evaluated at the black hole horizon. 
This expansion turns out to converge quite rapidly even at
temperatures $T \simeq 1.5 T_c$, quite opposite to the conventional
QCD perturbation theory at high temperature~\cite{Kajantie:2002wa}. We have extended our analysis to other thermodynamic quantities computed in the string frame, in particular the spatial string tension and the vacuum expectation value of the Polyakov loop, and the agreement with lattice data is better in this case.

From our analysis we see that the gravity model cannot reproduce at
the same time lattice data of the equation of state of the gluon
plasma at very high temperatures, and close to the phase
transition. This means that fixing the gravity constant $G_5$ from the
ideal gas limit seems not to be consistent with thermodynamics close
to the phase transition.  In this sense, there is the possibility that
the ideal gas limit doesn't correspond to the limit of the black hole
gravity theory at high temperatures. Is it possible that the gravity
theory allows more degrees of freedom at high temperatures?  Or is the
simulation of higher terms in the string coupling $\alpha'$ in the
gravity action incomplete?  Since the agreement of the velocity of
sound, the spatial string tension and the Polyakov loop in the string
frame is better, the question arises whether the gravity action
approximates the string action truthfully.  We will further address
this problem, and analyze possible solutions~\cite{Megias:pr} in forthcoming work.

The approach presented here and in previous references, see
e.g. Refs.~\cite{Gursoy:2008za,Alanen:2009xs}, is a
phenomenological gravity theory motivated by non-critical string
theory. The results are subject to $O(1)$ $\alpha^\prime$ corrections
and one can only hope that they capture the expected $\beta$-function
behavior. It has to be mentioned also as a caveat that the
contribution of the coupling of the world-sheet to the dilaton field
may very well change the quantitative, and even the qualitative result
substantially. This should be checked in future works.

\vspace{1cm}

{\bf Acknowledgments:}

E.M. would like to thank the Humboldt Foundation for their stipend. This work was also supported in part by the ExtreMe Matter Institute EMMI in the framework of the Helmholtz Alliance Program of the Helmholtz Association. We thank D.~Antonov, R.D.~Pisarski, E.~Ruiz Arriola and A.~Vairo for useful comments on the manuscript.

\newpage

\appendix{\textsf{\large Appendix A: Numerical solution of Einstein equations for the black hole metric}}
\label{sec:apA}
 
In this appendix we discuss in details the procedure to solve
numerically the system of Einstein equations given by
Eqs.~(\ref{eq:em1})-(\ref{eq:em4}). A numerical solution of the system
demands a good starting point. The boundary at $z=0$ has the
disadvantage that $b(z)$ is singular at this point.  The horizon at
$z=z_h$ is not a good expansion point either, since the inverse of the
black hole factor is singular there.  Practically it is possible to
start at some value close to the horizon, $z_i = z_h - \epsilon$. The
initial values can then be expanded in terms of their values at the
horizon as
\begin{eqnarray}
&&W(z_i) = W_h - W_h^\prime \,\epsilon +\frac{1}{2} W_h^{\prime\prime}
\,\epsilon^2 + \dots \,, \label{eq:Wi} \\ 
&&b(z_i) = b_h +
\frac{4}{9} b_h^2 W_h \,\epsilon + \frac{2}{9}\left(\frac{8}{9}
b_h W_h^2 - W_h^\prime \right) b_h^2 \,\epsilon^2 + \dots \,,
\label{eq:bi} \\ 
&&\alpha(z_i) = \alpha_h \Bigg[1 - \sqrt{b_h
W_h^\prime}\, \epsilon \label{eq:lambdai} +
\frac{1}{2}\left( b_h W_h^\prime -\frac{2}{9} b_h^\frac{3}{2}
W_h \sqrt{W_h^\prime} +\frac{1}{2}W_h^{\prime\prime} \sqrt{\frac{b_h}{W_h^\prime}} \right) \, \epsilon^2 + \cdots \Bigg] \,, \\ 
&& f(z_i)= 4 \pi T \, \epsilon + \frac{1}{2} b_h^2 V_h \,
\epsilon^2 + \frac{8}{9} \pi T \left( \frac{8}{9} b_h W_h^2 +
W_h^\prime \right) b_h \, \epsilon^3 + \dots \,, \label{eq:fi} \\ 
&&f'(z_i) = - 4 \pi T - b_h^2 V_h \, \epsilon -
\frac{8}{3} \pi T \left( \frac{8}{9} b_h W_h^2 + W_h^\prime \right)
b_h \, \epsilon^2 + \dots \,, \label{eq:fpi}
\end{eqnarray} 
where we use the notation
\begin{equation}
b_h \equiv b(z_h) \,,  \quad \alpha_h \equiv \alpha(z_h) \,, \quad W_h = W(z_h) \,, \quad V_h \equiv V(\alpha_h) \,.
\end{equation}
From Eqs.~(\ref{eq:fi}), (\ref{eq:fpi}) and $f(z_h) = 0$ one can derive
$W_h$ and its derivatives. The expression for $W_h^\prime$ follows
from Eq.~(\ref{eq:em1}) by applying l'H\^opital rule in the second term of the r.h.s. To compute the expression
of the second derivative $W_h^{\prime\prime}$, one derives Eq.~(\ref{eq:em1}) with respect to $z$ once, and uses the result of $W_h^\prime$. The result is
\begin{eqnarray}
&&W_h = - \frac{3}{16 \pi T} b_h V_h \,, \label{eq:Wh}\\ 
&&W_h^\prime =\left( \frac{3}{32 \pi T} \right)^2 b_h^3 \alpha_h^2 {\dot V}_h^2  \,, \label{eq:Wph} \\
&&W_h^{\prime\prime} =  -\left(\frac{b_h}{32\pi T}\right)^3 \Bigg[ 27 \alpha_h^3 b_h^2 {\dot V}_h^2 ({\dot V}_h + {\ddot V}_h \alpha_h) -60 \alpha_h^2 b_h^2 {\dot V}_h^2 V_h  + 256 b_h^2 V_h^3 \Bigg] \,,
\end{eqnarray}
where we have defined for simplicity of notation

\begin{equation}
{\dot V}_h \equiv \frac{dV}{d\alpha} \Bigg|_{\alpha_h} \,, \qquad 
{\ddot V}_h \equiv \frac{d^2 V}{d\alpha^2} \Bigg|_{\alpha_h} \,,
\end{equation}
and
\begin{equation}
V_h^\prime \equiv \frac{dV}{dz} \Bigg|_{z_h} =  -\frac{3}{32\pi T} \alpha_h^2 b_h^2 {\dot V}_h^2 \,.
\end{equation}
The derivative of $V$ with respect to~$\alpha$, i.e. ${\dot
V}(\alpha)$, can be computed analytically from the analytical
expression of $V(\alpha)$, c.f. Eq.~(\ref{eq:Vanalytic}).

Our procedure to solve the system of first order differential
equations Eqs.~(\ref{eq:em1})-(\ref{eq:em4}) follows
Refs.~\cite{Gursoy:2009jd,Alanen:2010tg}. First we choose arbitrary
values for the functions at the horizon, namely

\begin{equation}
b(\xi_h)=10 \,, \qquad \alpha(\xi_h)=0.5 \,,
\end{equation}
 and the initial values for temperature and $\epsilon$,
\begin{equation}
{\cal T} =  1 \, \textrm{GeV} \,, \qquad \epsilon = 10^{-9} \,\textrm{GeV}^{-1} \,.
\end{equation}
We rewrite the initial conditions, Eqs.~(\ref{eq:Wi})-(\ref{eq:fpi}),
in the $\xi$ coordinate, so that $\xi_i=0$ and $\xi_h = \xi_i +
\epsilon = \epsilon$. The parameter $\epsilon$ is chosen very small,
such that $\xi_i$ is very close to $\xi_h$ and the initial conditions
are accurate enough. The variables $\xi_h$, ${\cal T}$ differ from
$z_h$, $T$ by a rescaling factor.  Then one integrates numerically the
system to get solutions $W_1(\xi)$, $b_1(\xi)$, $\alpha_1(\xi)$ and
$f_1(\xi)$ in some interval $\xi_1 < \xi < \xi_h$, where $b_1$
diverges at $\xi_1$.  Since the system of
equations~(\ref{eq:em1})-(\ref{eq:em4}) is invariant under three
different rescalings~\cite{Gursoy:2009jd}, one can make use of these
properties to find a solution which has the right boundary
conditions. In step 2, one shifts the $\xi$ coordinate, so that the
ultraviolet divergence of $b$ is at the origin. The new solution reads

\begin{equation}
W_2(\rho) = W_1(\rho+\xi_1)  \,, \qquad b_2(\rho) = b_1(\rho+\xi_1)  \,, 
\qquad \alpha_2(\rho) = \alpha_1(\rho+\xi_1)  \,, \qquad f_2(\rho) = f_1(\rho+\xi_1) \,. \label{eq:sc1}
\end{equation}

In step 3, one chooses $\delta_f$ in such a way that $f_3(\rho = 0) = 1$, 
which is the correct value of $f$ in the UV,

\begin{equation}
W_3(\rho) = W_2(\rho) \sqrt{\delta_f} \,, \qquad b_3(\rho) =
b_2(\rho)/\sqrt{\delta_f} \,, \qquad \alpha_3(\rho) = \alpha_2(\rho) \,, \qquad
f_3(\rho) = f_2(\rho)/\delta_f \,. \label{eq:sc2}
\end{equation}

Finally one rescales $b_3$:

\begin{equation}
W(z) = W_3(z\delta_b) \,, \qquad b(z) = b_3(z\delta_b) \delta_b \,,
\qquad \alpha(z) = \alpha_3(z\delta_b) \,, \qquad f(z) =
f_3(z\delta_b) \,. \label{eq:sc3}
\end{equation}

The value $\delta_b$ is determined by comparing the zero temperature 
metric $b_0 (z_0)$ and the
finite temperature metric  $b_3 (\hat z)$ at a point $\hat z$ which has the same ultraviolet coupling 
as the zero temperature solution. In practice we
choose $\alpha_0(z_0)=0.07$. Then the rescaling factor $\delta_b$ is given as

\begin{eqnarray}
1        &=& \alpha_0(z_0)/\alpha_3 (\hat z)\\
\delta_b &=& b_0(z_0)/ b_3 (\hat z)\\
\hat z   &=& z_0 \delta_b.
\end{eqnarray}

Starting from the values of $b(\xi_h)$, $\alpha(\xi_h)$ and ${\cal T}$ mentioned above, one gets

\begin{equation}
\xi_1 = -0.334\, \textrm{GeV}^{-1} \,, \qquad
\delta_f = 0.960 \,, \qquad
\delta_b = 0.353 \,. \label{eq:xi1dfdb}
\end{equation}

By this procedure the QCD-parameter $\Lambda$ in the asymptotic logarithms
of both solutions also agree. 
Because $f'(z_h) = -4\pi T$, the operations performed in
Eqs.~(\ref{eq:sc1})-(\ref{eq:sc3}) rescales the value of $\cal T$ 
to the right temperature T,

\begin{eqnarray}
T   &=& \frac{\delta_b}{\delta_f}\cal{T} \,.  \label{eq:Trescaling}
\end{eqnarray}
From Eqs.~(\ref{eq:xi1dfdb}), (\ref{eq:Trescaling}) and the value
${\cal T} = 1 \,\textrm{GeV}$ one gets $T=368\, \textrm{MeV}$. In practice we solve the equations of motion for different temperatures by choosing
different values for $\alpha(\xi_h)$. Note that $\alpha(\xi_h)$ is
invariant under the set of rescaling Eqs.~(\ref{eq:sc1})-(\ref{eq:sc3}).


\vspace{1cm}

\appendix{\textsf{\large Appendix B: Ultraviolet properties of the thermodynamic quantities}}
\label{sec:apB}

We will study in this appendix the analytical properties of the thermodynamic quantities in the ultraviolet.~\footnote{In this analysis we will explicitly show every expression including all the perturbative orders one needs to compute the thermodynamic quantities up to ${\cal O}(\alpha^3)$.  } Being the UV expansion of $b(\alpha)$ given by (c.f.~Eq.~(\ref{eq:b0uv1}))
\begin{equation}
b(\alpha) = \frac{\ell}{z(\alpha)} \left[  1 -\frac{4}{9} \beta_0 \alpha +\frac{2}{81}\left(22\beta_0^2 -9\beta_1 \right) \alpha^2 
- \frac{4}{2187}\left(602\beta_0^3 - 540\beta_0\beta_1 + 81\beta_2 \right)\alpha^3
+ {\cal O}(\alpha^4) \right] \,, \label{eq:bzuv}
\end{equation}
the entropy can be computed by evaluating the above expression at the horizon, i.e. by computing $b(\alpha_h)$. The information on the horizon is given by the function $f(z)$, and so one should study its behavior in the UV. Combining Eqs.~(\ref{eq:em2}) and (\ref{eq:em4}) one gets
\begin{equation}
\frac{f''}{f'} + 3\frac{b'}{b} = 0 \,. \label{eq:ffbb}
\end{equation}
Given $b(z)$ one can solve this equation to get $f(z)$. Two integration constants are needed, which as usual are chosen by imposing the boundary conditions~$f(0)=1$ and $f(z_h)=0$~\cite{Gursoy:2008za}. The result for the UV asymptotics of $f(z)$ is
\begin{equation}
f(z) = 1 - \frac{z^4}{z_h^4} \frac{Q_f(z)}{Q_f(z_h)} \,, \label{eq:fzaz}
\end{equation}
where
\begin{equation}
Q_f(z) = 1+\frac{4}{3}\beta_0 \alpha(z) - \frac{1}{9}(7\beta_0^2-6\beta_1)\alpha^2(z) + \frac{1}{162}(271\beta_0^3-396\beta_0\beta_1 +72\beta_2)\alpha^3(z)  + {\cal O}(\alpha^4) \,, \label{eq:Cfz}
\end{equation}
and $\alpha(z)$ is given by Eq.~(\ref{eq:luv}). One can arrive at this result by considering the explicit expression of $b(z):=b(\alpha(z))$ given by Eq.~(\ref{eq:bzuv}), insert it into Eq.~(\ref{eq:ffbb}) and solve the equation reexpressing the result in powers $\alpha(z)$. A much easier way to arrive at this result is to consider a general scheme $f(z) = 1-(z^4/C_h) \cdot \left(1 + f_1\alpha(z) + f_2 \alpha^2(z) + f_3\alpha^3(z) + \dots \right)$, and then introduce it and Eq.~(\ref{eq:bzuv}) into Eq.~(\ref{eq:ffbb}). The derivative of $\alpha(z)$ is given by
\begin{equation}
z\frac{d\alpha}{dz} = \beta_0\alpha^2 + \beta_1\alpha^3 +\left( \frac{4}{9}\beta_0^3 + \beta_2 \right)\alpha^4 + {\cal O}(\alpha^5) \,. \label{eq:zdadz}
\end{equation}
This useful relation can be proved easily from the explicit expression of $\alpha(z)$, c.f.~Eq.~(\ref{eq:luv}). Using Eq.~(\ref{eq:zdadz}), the equation of motion Eq.~(\ref{eq:ffbb}) can be expressed in powers of $\alpha(z)$, and one can easily identify the coeficients $C_h$, $f_1$, $f_2$, $f_3, \dots$, which fulfill the equation.  The result is given by Eqs.~(\ref{eq:fzaz}) and (\ref{eq:Cfz}). 

From Eq.~(\ref{eq:fzaz}) and using Eq.~(\ref{eq:zdadz}), the derivative of $f(z)$ then evaluates to
\begin{eqnarray}
f'(z) &=& \frac{-4 z^3}{z_h^4 Q_f(z_h)} \Bigg(1+\frac{4}{3}\beta_0 \alpha(z) - \frac{2}{9}(2\beta_0^2-3\beta_1)\alpha^2(z)  \nonumber \\
&&\qquad\qquad\qquad\qquad + \frac{4}{81}(26\beta_0^3-36\beta_0\beta_1 + 9\beta_2)\alpha^3(z)
 + {\cal O}(\alpha^4)\Bigg) \,.
\end{eqnarray}
The temperature is obtained by evaluating the above expression at the horizon
\begin{equation}
T = -\frac{f'(z_h)}{4\pi} = \frac{1}{\pi z_h} \left[ 1 +\frac{\beta_0^2}{3} \alpha_h^2  -\frac{\beta_0}{6}(5\beta_0^2-4\beta_1)\alpha_h^3   + {\cal O}(\alpha_h^4)  \right] \,. \label{eq:Tzhuv}
\end{equation}
We have corrected some error in the computation of Ref.~\cite{Gursoy:2008za}, where the authors consider a factor $-4/9$ instead of $1/3$ at order ${\cal O}(\alpha_h^2)$ in the bracket of Eq.~(\ref{eq:Tzhuv}). 

The entropy density easily follows by evaluating Eq.~(\ref{eq:bzuv}) at the horizon, and using the relation between $z_h$ and $T$ given by Eq.~(\ref{eq:Tzhuv}). Then one gets
\begin{eqnarray}
s(T) &=& \frac{1}{4G_5} b^3(z_h)  = \frac{\pi^3 \ell^3}{4 G_5} T^3  \Bigg[ 1 -\frac{4}{3} \beta_0 \alpha_h + \frac{1}{9} \left( 11\beta_0^2 -6\beta_1 \right) \alpha_h^2  \nonumber \\
&&\qquad\qquad\qquad\qquad\qquad -\frac{1}{162}\left(163\beta_0^3 -252\beta_0\beta_1 +72\beta_2 \right) \alpha_h^3
+ {\cal O}(\alpha_h^4) \Bigg] \,, \label{eq:s1}
\end{eqnarray}
which corresponds to the weak coupling expansion of the entropy. To deal with Eq.~(\ref{eq:s1}) one needs to know the temperature dependence of $\alpha_h$. Taking into account the relation between $T$ and $z_h$ given by Eq.~(\ref{eq:Tzhuv}), it can be proved that~\footnote{Note that because of Eq.~(\ref{eq:ahaapp}) the expressions of the thermodynamics quantities up to ${\cal O}(\alpha_h^3)$ remain valid when one substitutes $\alpha_h$ by $\alpha_T$.}
\begin{equation}
\alpha_h = \alpha_T + \frac{\beta_0^3}{3}\alpha_T^4 + {\cal O}(\alpha_T^5) \,, \qquad \alpha_T \equiv \alpha\left(z=\frac{1}{\pi T}\right) \,, \label{eq:ahaapp}
\end{equation}
where $\alpha_T$ is defined as indicated. One can prove that by using the explicit expansion given by Eq.~(\ref{eq:luv}). By considering $z\to 1/(\pi T)$ in Eq.~(\ref{eq:zdadz}) one gets~\footnote{The definition of $\alpha_T$ given in Eq.~(\ref{eq:ahaapp}) doesn't agree with the usual definition of the running coupling at finite temperature, for which the prescription $E\simeq \pi T$ is usually taken. Both prescriptions differ at ${\cal O}(\alpha_T^3)$,  i.e. $\alpha_E = \alpha_T + {\cal O}(\alpha_T^3)$,  as it is evident from Eq.~(\ref{eq:dahdtex}).}
\begin{equation}
T\frac{d\alpha_{T,h}}{dT} = -\beta_0\alpha_{T,h}^2 -\beta_1 \alpha_{T,h}^3 - \left( \frac{4}{9}\beta_0^3 + \beta_2 \right)\alpha_{T,h}^4 + {\cal O}(\alpha_{T,h}^5) \,. \label{eq:dahdtex}
\end{equation}
The $\alpha_h$ version of this formula easily follows by considering Eq.~(\ref{eq:ahaapp}). Eq.~(\ref{eq:dahdtex}) is very useful, and it can be used for instance to compute easily the pressure from Eq.~(\ref{eq:s1}) (see Section~\ref{sec:Bekenstein_Hawking} for a discussion). The energy density follows trivially from Eqs.~(\ref{eq:s1}) and (\ref{eq:pwc})
\begin{eqnarray}
\frac{\epsilon(T)}{T^4} = \frac{s}{T^3} - \frac{p}{T^4} &=& \frac{3\pi^3 \ell^3}{16 G_5} \Bigg[ 1 - \frac{4}{3}\beta_0\alpha_h + \frac{2}{3}(2\beta_0^2 - \beta_1) \alpha_h^2  \nonumber \\
&&\qquad\qquad\quad -\frac{1}{162}\left( 187\beta_0^3 -288\beta_0\beta_1 +72\beta_2 \right) \alpha_h^3
+ {\cal O}(\alpha_h^4)  \Bigg] \,. \label{eq:ewc} 
\end{eqnarray}
Note that in the weak coupling expansion of the thermodynamics quantities, Eqs.~(\ref{eq:s1}), (\ref{eq:pwc}) and (\ref{eq:ewc}), there are no half-integer powers in $\alpha$, i.e. $\alpha^{3/2}$, $\alpha^{5/2}, \dots\,$, as we don't consider loops contributions, in contrast to the weak coupling expansion in QCD~\cite{Kajantie:2002wa}.  These results extend to ${\cal O}(\alpha_h^3)$ the results of Ref.~\cite{Gursoy:2008za}. From the energy density and pressure one can compute the trace anomaly. It reads

\begin{equation}
\Delta(T)=  \frac{\beta(\alpha)}{4\alpha^2} \frac{\langle \textrm{Tr} F_{\mu\nu}^2 \rangle}{T^4}  = \frac{\epsilon-3p}{T^4} = \frac{\pi^3\ell^3}{12 G_5} \Bigg[ 
\beta_0^2 \alpha_h^2 
-\frac{2}{3}\beta_0\left(2\beta_0^2-3\beta_1\right)\alpha_h^3
+ {\cal O}(\alpha_h^4) \Bigg] \,. \label{eq:deltauv} 
\end{equation}
The trace anomaly is related to the vacuum expectation value of the gluon condensate. As it was pointed out in Ref.~\cite{Gursoy:2008za} and discussed by us in Sec.~\ref{sec:bhsol}, the gluon condensate appears in the UV expansion of the difference between the black-hole and zero temperature solutions, c.f. Eqs.~(\ref{eq:bz4})-(\ref{eq:fz4}). In this Appendix we have not taken into account power corrections in~$z$. However, just by computing $b^3(z_h)$ using Eq.~(\ref{eq:bz4}), it is straightforward to prove that the correction $\sim z^4$ induces a contribution $\sim 1/T^4$ in $s(T)/T^3$, $p(T)/T^4$, $\epsilon(T)/T^4$ and $\Delta(T)$, and so this corresponds to much a lower order contribution in the UV expansion performed previously. By the same way, the power correction in $\alpha(z)$ induces a correction $\sim 1/T^4$ in $\alpha(z_h)$ which is subleading in our UV analysis, and so it is enough to identify $\alpha(z_h)$ with $\alpha_0(z_h)$ at this level, as it has been done in previous formulas. 

From previous analysis we can derive easily the expressions for the weak coupling expansion of the specific heat per unit volume
\begin{eqnarray}
c_v &=& T \frac{\partial^2 p}{\partial T^2} = \frac{3\pi^3\ell^3}{4 G_5} T^3 \Bigg[ 
 1 - \frac{4}{3}\beta_0 \alpha_h + \frac{1}{3}\left( 5\beta_0^2-2\beta_1\right)\alpha_h^2  \nonumber \\
&&\qquad\qquad\qquad\qquad\qquad + \frac{1}{162}\left( -295\beta_0^3 + 396\beta_0\beta_1 - 72\beta_2 \right)\alpha_h^3 + {\cal O}(\alpha_h^4)
\Bigg] \,, \label{eq:cv}
\end{eqnarray}
and speed of sound
\begin{eqnarray}
c_s^2 &=& \frac{s}{c_v} = \frac{1}{3} \Bigg[ 
 1 - \frac{4}{9}\beta_0^2 \alpha_h^2 + \frac{2}{9}\beta_0\left( \beta_0^2-4\beta_1\right)\alpha_h^3 + {\cal O}(\alpha_h^4)
\Bigg] \,. \label{eq:cs}
\end{eqnarray}
Eq.~(\ref{eq:cv}) follows by using Eqs.~(\ref{eq:dahdtex}) and (\ref{eq:pwc}), while Eq.~(\ref{eq:cs}) is obtained from Eqs.~(\ref{eq:s1}) and (\ref{eq:cv}).

For completeness of this apendix, we study the high temperature behavior of the Polyakov loop. At the end one wants to express the result as an expansion in powers of $\alpha_h$, and the easiest way to proceed is to work in coordinates dependent on the running coupling~$\alpha$ as a variable, instead of $z$. The relation between $z$ and $\alpha$ is given by~\cite{Galow:2009kw}
\begin{equation}
\frac{d\alpha}{dz} = \frac{1}{\ell} b(\alpha) e^{-D(\alpha)} \,, \label{eq:dadzD}
\end{equation}
where the function $e^{D(\alpha)}$ reads
\begin{equation}
e^{D(\alpha)} = -\frac{1}{\beta(\alpha)} \exp \left[\frac{4}{3} \int_0^\alpha \frac{\beta(a)}{3a^2} da \right] \,. \label{eq:defeD}
\end{equation}
Using Eqs.~(\ref{eq:dadzD}), one can rewrite the Nambu-Goto action which involves an integration in~$z$, Eq.~(\ref{eq:NG_action2}), as
\begin{equation}
S_{\textrm{\tiny NG}} = \frac{\ell}{2\pi l_s^2 T} \int_0^{\alpha_h} da \, e^{D(a)} b(a) a^{\frac{4}{3}} \,. \label{eq:NG_action3}
\end{equation}
In the intermediate steps we will make use explicitly of the UV $\beta$-function up to 4-loops order just for completeness, i.e.
\begin{equation}
\beta(\alpha) = -\beta_0 \alpha^2 - \beta_1 \alpha^3 - \beta_2 \alpha^4 -\beta_3 \alpha^5 + \cdots \,,\label{eq:betauv4loop}
\end{equation}
but our final result of the Polyakov loop up to ${\cal O}(\alpha_h^{10/3})$ will depend only on $\beta_0$ and $\beta_1$, c.f. Eq.~(\ref{eq:L3uv}). Inserting the UV $\beta$-function, Eq.~(\ref{eq:betauv4loop}), into Eq.~(\ref{eq:defeD}), one gets
\begin{equation}
e^{D(\alpha)} = \frac{1}{\beta_0 \alpha^2} - \left( \frac{4}{9} + \frac{\beta_1}{\beta_0^2} \right)\frac{1}{\alpha} + \frac{1}{81}\left( 8\beta_0 
 + 18\frac{\beta_1}{\beta_0} +\frac{\beta_1^2}{\beta_0^3} -\frac{\beta_2}{\beta_0^2} \right)  +  {\cal O}(\alpha) \,. \label{eq:eDuv}
\end{equation}
Note that Eq.~(\ref{eq:dadzD}) combined with the expansions of $b(\alpha)$ and $e^{D(\alpha)}$ given by Eqs.~(\ref{eq:bzuv}) and (\ref{eq:eDuv}) respectively, leads to Eq.~(\ref{eq:zdadz}). The main difficulty is to express $b$ as a function of $\alpha$. The UV expansion of $b(\alpha)$ is given by Eq.~(\ref{eq:bzuv}). To compute $z(\alpha)$ one has to invert Eq.~(\ref{eq:luv}). The result is
\begin{eqnarray}
z &=&  \frac{1}{\Lambda (\beta_0\alpha)^{\beta_1/\beta_0^2}} \exp\Bigg[ -\frac{1}{\beta_0\alpha} -\beta_0 K + \left( -\frac{4}{9}\beta_0 + \frac{\beta_1^2}{\beta_0^3} - \frac{\beta_2}{\beta_0^2} \right)\alpha  \nonumber \\
&&\qquad + \left(\frac{4}{9}\beta_0^2 -\frac{2}{9}\beta_1 - \frac{1}{2\beta_0^4}\left( \beta_1^3 -2\beta_0\beta_1\beta_2 + \beta_0^2\beta_3 \right) \right)\alpha^2   + {\cal O}(\alpha^3) \Bigg] \,.  \label{eq:zuv}
\end{eqnarray}
From Eqs.~(\ref{eq:bzuv}) and (\ref{eq:zuv}), one gets
\begin{eqnarray}
b(\alpha) &=& \ell\Lambda (\beta_0\alpha)^\frac{\beta_1}{\beta_0^2} e^{\frac{1}{\beta_0\alpha} + \beta_0 K} \cdot \Bigg[ 1 + \frac{\beta_0\beta_2 - \beta_1^2}{\beta_0^3}\alpha  \nonumber \\
&&\qquad + \frac{1}{2\beta_0^6}\left(\beta_1^4-2\beta_0\beta_1\beta_2(\beta_0^2+\beta_1) +\beta_0^2(\beta_1^3+\beta_2^2 + \beta_0^2\beta_3)  \right)\alpha^2  
+ {\cal O}(\alpha^3)   \Bigg] \,. \label{eq:bauv}
\end{eqnarray}
Then inserting Eqs.~(\ref{eq:eDuv}) and (\ref{eq:bauv}) into Eq.~(\ref{eq:NG_action3}), and performing the integration in $a$, one arrives at
\begin{equation}
S_{\textrm{\tiny NG}} = S^0_{\textrm{\tiny NG}} -\frac{\ell^2}{2 l_s^2} \alpha_h^{\frac{4}{3}} \left( 1 + \frac{4}{9}\beta_0\alpha_h + \frac{1}{81}\left(161\beta_0^2+72\beta_1 \right)\alpha_h^2 + {\cal O}(\alpha_h^3) \right) \,, \label{eq:SngL1}
\end{equation}
where $S^0_{\textrm{\tiny NG}}$ is divergent and comes from the lower limit in the integration. To arrive at Eq.~(\ref{eq:SngL1}) one has to make use of Eq.~(\ref{eq:zuv}) evaluated at the horizon, and use the relation between $T$ and $z_h$ given by Eq.~(\ref{eq:Tzhuv}). Then the renormalized vacuum expectation value of the Polyakov loop writes
\begin{equation}
L_R(T) = e^{-S^{\textrm{\tiny reg}}_{\textrm{\tiny NG}}} = \exp\left[ \frac{\ell^2}{2 l_s^2} \alpha_h^{\frac{4}{3}} \left( 1 + \frac{4}{9}\beta_0\alpha_h + \frac{1}{81}\left(161\beta_0^2+72\beta_1 \right)\alpha_h^2  + {\cal O}(\alpha_h^3) \right)\right] \,, \label{eq:L3uv}
\end{equation}
where $S^{\textrm{\tiny reg}}_{\textrm{\tiny NG}} \equiv S_{\textrm{\tiny NG}} - S^0_{\textrm{\tiny NG}}$. Note that $L_R(T)$ tends to $1$ from above in the high temperature limit, which is the behavior predicted by standard perturbative QCD.

\bibliographystyle{h-physrev3}
\bibliography{Refs}

\end{document}